\documentclass[preprint]{aastex}
\usepackage[]{natbib}
\usepackage[]{verbatim}
\usepackage{algorithm}
\usepackage{algorithmic}
\usepackage{amsmath}
\usepackage{graphicx}
\usepackage{etex}
\reserveinserts{18}
\usepackage{morefloats}
\usepackage{lscape}
\usepackage{color}
\usepackage{ulem}
\usepackage{amssymb}

\newcommand{\kms}{km\ s$^{-1}$}

\newcommand{\gsim}{\stackrel{\scriptscriptstyle >}{\scriptstyle {}_\sim}}

\newcommand{\ha}{H$\alpha$}

\newcommand{\hb}{H$\beta$}
\newcommand{\oiii}{[O\,{\sc iii}]}

\newcommand{\civ}{C\,{\sc iv}}

\newcommand{\mgii}{Mg\,{\sc ii}}

\newcommand{\lam}{$\lambda$}
\newcommand{\Msigma}{\ifmmode M_{\rm BH} - \sigma \else $M_{\rm BH} - \sigma$\fi}

\newcommand{\nii}{[N\,{\sc ii}]}

\newcommand{\nev}{[Ne\,{\sc v}]}
\newcommand{\Halpha}{\ifmmode {\rm H}\alpha \else H$\alpha$\fi}

\newcommand{\Hbeta}{\ifmmode {\rm H}\beta \else H$\beta$\fi}

\newcommand{\spitzer}{{\it Spitzer}}
\newcommand{\herschel}{{\it Herschel}}
\newcommand{\galex}{\textit{GALEX}}
\newcommand{\wise}{{\it WISE}}

\shorttitle{AGN LoCuSS}
\begin{document}

\title{A Herschel Study of 24 $\mu$m-Selected AGNs and Their Host Galaxies}

\author{Xu, Lei\altaffilmark{1}, Rieke, G. H.\altaffilmark{1}, Egami, E.\altaffilmark{1},  Pereira, M. J.\altaffilmark{1}, Haines, C. P.\altaffilmark{1,3}, \& Smith, G. P.\altaffilmark{2}}

\altaffiltext{1}{ Steward Observatory, 933 N. Cherry Ave, University of Arizona, Tucson, AZ 85721, USA}
\altaffiltext{2}{School of Physics and Astronomy, University of Birmingham, Birmingham B15 2TT, UK }
\altaffiltext{3}{Departamento de Astronomìa, Universidad de Chile, Casilla 36-D, Correo Central, Santiago, Chile}


\begin{abstract}
We present a sample of 290 24$\mu$m-selected active galactic nuclei (AGNs) mostly at $z \sim 0.3-2.5$, 
within 5.2 $\rm {deg}^2$ distributed as $25\rm'\times25\rm'$ fields
around each of 30 galaxy clusters in 
the Local Cluster Substructure Survey (LoCuSS). 
The sample is nearly complete to 1 mJy at 
24$\mu$m, and has a rich multi-wavelength set of ancillary data; 162 are detected by {\it Herschel}.
We use spectral templates for AGNs, stellar populations, and infrared emission by star forming galaxies 
to decompose the spectral energy distributions (SEDs) of these AGNs
and their host galaxies, and estimate their star formation rates (SFRs), 
AGN luminosities, and host galaxy stellar masses. The set of templates is relatively simple:  
a standard Type-1 quasar template; another for the photospheric output of the stellar population; and 
a far infrared star-formng template. For the Type-2 AGN SEDs, we substitute templates including 
internal obscuration, and some Type-1 objects require a  warm 
component ($T \gsim 50 $ K). 
The individually {\it Herschel}-detected Type-1 AGNs and a subset of 17 Type-2 ones 
typically have luminosities $ > 10^{45} \rm ~ergs~s^{-1}$, and supermassive black holes of 
$\sim 3 \times 10^8 M_\odot$ emitting at $\sim$ 10\% of the Eddington rate. We find them in  
about twice the numbers of AGN identified in SDSS data in the same fields, i.e., they represent 
typical high luminosity AGN, not an infrared-selected minority. These AGNs and their host galaxies are studied 
further in an accompanying paper.


\end{abstract}


\section{Introduction}
\label{section:intro}

The bright continua of  active galactic nuclei (AGNs) 
 in the X-ray, UV and optical are powered directly by accretion -- the growth of super massive black holes (SMBHs).
At the current epoch, there is a tight correlation between the SMBH masses 
and the host galaxy stellar bulge masses, indicating a link
between the integrated accretion by black holes and the star formation in their host galaxies
(e.g., Magorrian et al. 1998; Tremaine et al. 2002).
The level of accretion is indicated by a variety of metrics, e.g. X-rays,
optical emission lines, and optical-IR continua. These indicators can 
drown out many metrics for the level of star formation, but it is
thought that the far infrared (FIR) emission remains dominated by this mechanism, providing 
a strong motivation for studies of the FIR outputs of galaxies with active nuclei.

Prior to \herschel\/, the measurements of rest-frame FIR emission from luminous AGNs were limited to 
a small population (e.g.,  Omont et al. 2001; Haas et al. 2003; Dicken et al. 2008).
With the advent of \herschel\/, it is possible 
to study  the FIR properties efficiently for a large sample of  AGNs \citep[e.g.,][]{hatziminaoglou10,
shao10, mullaney12, rosario13, leipski13,leipski14}. To augment these studies, we describe {\it Herschel} measurements of 
205 Type 1 AGNs uniformly selected from a 5.2 $\rm {deg}^2$ survey area;
the sample is nearly complete at   24 $\mu$m down to 1 mJy, and verified by spectroscopy. 
These AGNs are complemented by 85 Type 2 objects similarly selected from a 3.6  $\rm {deg}^2$ 
subset of the same data. 
A multi-band dataset from the UV to the FIR, including optical spectroscopy, 
allows detailed study of the AGNs and their host galaxies in this sample.
We use spectral templates for 1.) the UV to far infrared output of AGNs; 2.) stellar populations; and 
3.) the infrared emission of star forming galaxies 
to decompose their spectral energy distributions (SEDs) and  
 estimate their IR star formation luminosities,
AGN luminosities, and their host galaxy stellar masses.
The black hole masses are estimated from the Type-1 AGN broad optical emission lines, and 
from the stellar masses for the Type-2 objects. 
More than 55\% of the sample are detected individually by {\it Herschel}, and
we stack the signals from the rest for comparison purposes.
This paper presents the data and a basic analysis; in a companion paper 
(Xu et al. 2015), we use these results to explore the evolutionary stage of the AGN-host galaxies --
whether they are starbursting or normal star-forming galaxies; and whether there is 
a causal connection between nuclear and SF activity for these objects.
 
This paper is structured as follows. In Section 2
we present the data, and in Section 3 we describe the selection and completeness of our sample of Type-1 AGNs. 
We also show their SEDs and discuss the FIR excesses.  Section 4 is a parallel discussion of the Type-2 objects.
In Section 5 we analyze both samples together, calculating the physical properties of the AGNs and their 
host galaxies, such as Eddington rates, host stellar masses and star formation rates. A summary is provided in Section 6.  
Throughout this paper we 
assume $\Omega_{M}$\,=\,0.3, $\Omega_{\Lambda}$\,=\,0.7, and $H_0$\,=\,70\,km\,s$^{-1}$\,Mpc$^{-1}$.

\section{DATA\label{data_type1}}

\subsection{LIRAS: LoCuSS Infrared AGN Survey}

The {\bf Lo}cal {\bf C}l{\bf u}ster {\bf S}ubstructure {\bf S}urvey
(LoCuSS)\footnote{http://www.sr.bham.ac.uk/locuss/}
is a large survey of X-ray luminous galaxy clusters at
$z=0.15-0.3$ (e.g., Smith et al. 2010).  This paper exploits the extensive LoCuSS
multi-wavelength data set for 30
clusters, which includes data from $Chandra$, $GALEX$, SUBARU, UKIRT,
$Spitzer$/MIPS, and \herschel.  The $Spitzer$ and \herschel\/ data
cover a total area of $\sim$ 5.2 deg$^{2}$ ($25\arcmin \times 25\arcmin \times 30$), 
at the central coordinates listed in Table \ref{cluster}. 
Since most cluster members are members of the old galaxy population,
which is not bright in the MIR and FIR, 
this wide-field coverage allows us to conduct a serendipitous
AGN survey independent of the existence of galaxy clusters in the observed
fields. In LIRAS  ({\bf L}ocuSS {\bf I}nfra{\bf r}ed {\bf A}GN {\bf S}urvey), 
we take advantage of these multi-wavelength datasets (described in this section) to study the 
properties of a 24 $\mu$m-selected IR luminous Type-1 AGN sample over the entire  5.2 deg$^{2}$ area (as discussed 
in Section 3 below). 
In Section 4, we show how we also identified Type-2 AGNs selected from 21 out of the 30 cluster 
fields (i.e., 3.6 $\rm {deg}^2$), as indicated in Table \ref{cluster}.


\subsection{Mid-infrared Observations}
\label{section:mips}

Each cluster field was observed at 24 $\mu$m  between November 2007 and November 2008 with MIPS \citep{rieke04} on 
$Spitzer$  
\citep{werner04}, utilizing
a $5\times5$ grid of pointings in fixed cluster or raster
mode (PID: 40872; PI: G.P. Smith). Two cycles of  small-field photometry
with a frame time of 3 s were performed at each grid point, for a total
per pixel exposure time of 90 s. The central $5\arcmin \times 5\arcmin$ 
of some clusters had already been imaged by GTO program 83 to greater
depth ($\sim 3000$ s/pixel). All the available data were combined for our survey.  
The images were processed with the MIPS Data Analysis Tool (DAT; Gordon et al. 2005).
The beam size at 24 $\mu$m  is 5$\farcs$9 with 2$\farcs$49 pixels; the images 
were combined with a pixel scale of
 $1\farcs245$, half the physical pixel scale.
The 24 $\mu$m fluxes were measured by SExtractor \citep{bertin96}
 within a fixed circular aperture of diameter 21$\arcsec$ and with an 
aperture correction of a factor 1.29.
 The 90\% completeness limits at 24 $\mu$m
are in the range 300-500 $\mu$Jy.
Details of the reduction, source extraction and photometry can
be found in \citet{haines09a}.

WISE\footnote{http://irsa.ipac.caltech.edu/Missions/wise.html} 
data are also available in our survey fields.  
We utilize WISE 3.4 $\mu$m, 4.6 $\mu$m, and 
12 $\mu$m  measurements in our SED decomposition fitting. The detection limit at 
22 $\mu$m is 6 mJy, an order of magnitude higher than achieved with MIPS.  
Since this band is  so close to the MIPS 24 $\mu$m one, 
we do not use it.

\subsection{ Far-infrared Data}
\label{section:herschel}

Our \herschel\/ (Pilbratt et al. 2010) data were taken  
between 22 December 2009 and 10 October 2011 (LoCuSS \herschel\/ key Programme, Smith et al. 2010).
Each cluster field was observed  
with both the Photodetector Array Camera and Spectrometer (PACS; Poglitsch et al. 2010) at 
100 and 160 $\mu$m, 
and the Spectral and Photometric
Imaging Receiver (SPIRE; Griffin et al. 2010) at  250, 350, and 500 $\mu$m.
The images were reduced using HIPE V6.0 (Ott 2010).
Because the PACS images are relatively shallow and the beam size is relatively small, 
confusion is not an issue and the photometry could be performed with SExtractor. 
However, confusion is an issue for the SPIRE data.  
Therefore the photometry of the SPIRE images was
performed with IRAF/DAOphot, using the 24 $\mu$m source
positions (for all sources above the 3$\sigma$ detection limit) as priors
to position the PSF (point spread function) on the SPIRE maps.
We rotated the {\it Herschel} PSF to match the position angle of each map,
registered the 24 $\mu$m and \herschel\/ maps with
the  isolated point sources, and then fixed 
the source positions. We extracted fluxes on the SPIRE maps using the empirical fine scale PSF provided by
the HSC \footnote{ftp://ftp.sciops.esa.int/pub/hsc-calibration/SPIRE/PHOT/Beams\_v1.0/}
instead of constructing one from our own data, because of the lack of 
isolated point sources with high ratios of signal to noise on our maps. 
Parameters adopted for  the maps and photometry 
(such as the pixel size, FWHM of point source, photometry aperture radius, 
aperture correction, and  sky annulus radius)
 are summarized in Table \ref{tab:herschel_photo}.

\subsection{Near-infrared, Optical, and Ultraviolet  Data}

Near-infrared images of 26 of the
30 cluster fields were obtained with WFCAM (Casali et al.
2007) at J- and K-bands on the 3.8-m United Kingdom Infrared Telescope
(UKIRT)
in service mode over multiple semesters starting
in March 2008. The data acquisition used 
the same strategy as was used by the UKIDSS
Deep Extragalactic Survey (Lawrence et al. 2007), covering
$52\arcmin \times 52\arcmin$  to depths of J $\sim$ 21,  K $\sim$ 19, with exposure
times of 640 s, pixel size of $0\farcs2$ (half the physical pixel size) and PSF FWHMs $\sim0\farcs7-1\farcs2$.
The remaining four cluster fields were observed with NEWFIRM
on the 4.0-m Mayall telescope at Kitt Peak
on 17 May 2008 and 28 December 2008. The NEWFIRM
data consist of dithered and stacked J- and K-band images
covering fields of $27\arcmin \times 27\arcmin$ with a $0\farcs4$
pixel-scale and PSF FWHM $\sim 1\farcs0-1\farcs5$. The total exposure
times in each filter were 1800 s, and the images also reach depths of
J$\sim$21 mag and K$\sim$19 mag (Vega, 5-sigma).

SDSS photometry (Data Release 7) is available for 26 out of 30 cluster fields, 
covering a total of 4.51 $\rm {deg}^2$ survey area.
The SDSS five band photometry we used is corrected for Galactic extinction.
Optical images in R or I band using Subaru/Suprime-Cam \citep{okabe10} with
seeing $\sim$ 0\farcs6 allow us to study the
morphology of the Type 2 AGN hosts. 
The data were reduced as described by \citet{okabe08}, using the Suprime-Cam pipeline
software SDFRED for flat-
fielding, instrumental distortion correction, differential refraction,
PSF matching, sky subtraction, and stacking. The astrometric solution was based on
2MASS stars. Standard stars were interspersed with the cluster imaging. Further
information about the optical imaging, including the initial publication of the data for most of the clusters,
can be found in \citet{okabe10}.

 GALEX NUV observations were obtained for 26 of the cluster fields (omitting those for
A586, A689, A2485, and RXJ0142), and simultaneously in the FUV for 21 fields under
Guest Investigator Programs GI4-090 and GI6-046 (PIs G. P. Smith and S. Moran,
respectively). The exposure times ranged from 3 to 29 ks; additional details about
these observations and their reduction can be found in \citet{haines2015}.


\subsection{$Chandra$ X-ray imaging}

Twenty one of the 30 clusters were observed with Chandra 
in the I mode of the Advanced
Camera for Imaging Spectroscopy (ACIS-I), which
has a field of view of $16\farcm9 \times 16\farcm9$. 
Seven more were observed with the ACIS-S ($8\farcm3 \times 8\farcm3$ FOV). 
Two of the clusters (Abell 2345 and Abell 291) do not have X-ray data.
The exposure times for the cluster fields range 
from 10 ks to 100 ks (Table 1 in Haines et al. 2012), with a
typical integration time of 20 ks.

Sanderson et al. (2009) discuss the reduction
and analysis of the X-ray observations. To detect X-ray point-sources that are potential AGNs, 
we used the wavelet-detection algorithm {\sc ciao wavdetect}; 
a minimum of six counts in the broad energy (0.3--7\,keV) 
range was the threshold for source detection.
The observations in this band were converted to fluxes assuming a $\Gamma{=}1.7$ 
power-law spectrum with Galactic absorption, following \citet{kenter05}. 
The X-ray flux sensitivity limit 
for the cluster fields ranges from $6\times 10^{-16} $ to $8\times 10^{-15}$  $\rm ergs~cm^{-2}~s^{-1}$
with a median of $3.5\times 10^{-15} $ $\rm ergs~cm^{-2}~s^{-1}$.
We calculated the luminosities for all sources with redshifts,
 assuming K-corrections of the form $(1+z)^{\Gamma-2}$.
We also calculated the X-ray luminosity limit for each non-detected AGN.


\subsection{Spectroscopic data}
We use spectra from ACReS (the Arizona Cluster
Redshift Survey; Pereira et al. 2015 in preparation) a
long-term spectroscopic program to observe the fields 
of the 30 galaxy clusters with MMT/Hectospec. Hectospec
is a 300-fiber multi-object spectrograph with a circular
field of view of $1\arcdeg$ diameter (Fabricant et al. 2005) 
and fibers that project to 1\farcs5 on the sky, mounted 
on the 6.5-m MMT at Mount Hopkins, Arizona.
We used the 270 line grating, which provides a wide wavelength
range (3650\AA--9200\AA) at 6.2 \AA~ resolution ($R=1000$). This ensures
coverage of the most important emission lines 
suitable
for identifying AGNs. 
The spectroscopy data were reduced using 
HSRED\footnote{http://mmto.org/~rcool/}.
Redshifts were determined by comparison of the reduced spectra
with stellar, galaxy and quasar template spectra, choosing
the template and redshift  that minimized the $\chi^2$
between model and data.
The target selection is described in detail in \citet{haines2013}. Virtually 
all sources with 24 $\mu$m flux above 1 mJy, and K-band $< 19$ mag, were
targeted by Hectospec.
See Section \ref{sample_type1}  and Appendix \ref{coverage} for a summary of the spectroscopic 
coverage of the 24 $\mu$m sources.

\section{24 $\mu$m-Selected Type-1 AGNs \label{sample_type1}}

The mid-IR continuum emission of  Type-1 AGNs  arises from warm dust heated by the AGN
 (e.g., Rieke 1978; Polletta et al. 2000;  Haas et al. 2003), and on average there are only modest variations among quasars 
in the average fraction of the bolometric luminosity emitted at these wavelengths \citep{krawczyk13}.
Therefore, the 24 $\mu$m selection of Type-1 AGN is expected to be highly
efficient and complete. 

\subsection{Type-1 AGN Identification}

\subsubsection{Approach}

Members of our sample of LIRAS Type-1 AGNs (see Table 3 for those detected and Table 4 for those undetected with {\it Herschel})
 were required to have:
\begin{enumerate}

  \item  $Spitzer$/MIPS 24 $\mu$m flux densities above 1 mJy; and

  \item  optical spectra showing broad emission lines with full width at half maximum (FWHM)
   $>1200$ \kms.

\end{enumerate}

All sources in the surveyed regions with 24 $\mu$m flux densities above 1 mJy and with K-band $< 19$ mag 
were given the  highest priority for spectroscopy, 
irrespective of near-IR color or morphology (resolved/unresolved
in the near-IR). We excluded those that were
clearly stars (being both unresolved in the K-band data, and having 
blue near-IR colors ($\rm J-K<1.0$)) and asteroids 
 (very luminous at 24 $\mu$m, but with no counterparts in all
other bands). Over the 5.2 $\rm {deg}^2$ survey area covered at both 24 $\mu$m
and in the  \herschel\/ bands, there were 2439 sources  with  24 $\mu$m flux density above 1 mJy, 
of which 1827 remained after excluding stars, asteroids, and sources with no optical/NIR counterparts. 
From this list, 1729 sources were observed with Hectospec while another 18 
have spectra from SDSS.  Therefore, the completeness
of spectroscopic coverage is about 94.6\% (See Appendix A for 
a summary of the reasons that we missed 5.4\%  of the targets). 
Among the 1729 sources targeted by Hectospec, 1263 
yielded spectroscopic redshifts with a corresponding success rate
of 73\%.  

To identify Type-1 AGNs, we
fitted each emission line  in the optical spectra with single or double Gaussian profiles. 
We list the FWHM of typical broad emission lines in  Table 5.
Sources showing emission
lines (specifically, \mgii, \civ, \Hbeta, or \Halpha) with FWHM over
1200 \kms~  were selected as Type-1 AGNs  \citep{hao05}.  Finally 205 sources satisfied our Type-1 AGN selection
criteria, 177 confirmed with Hectospec and 28 confirmed with SDSS\footnote{The spectra were
inspected visually to confirm the classification. One ambiguous case (\#6 J084352.28+292854.0)
was retained because its SED decomposition fit (see Section 3.4) supported the classification.
J084234.94+362503.2 (\#13) falls slightly below our line width criterion but was
retained because spectropolarimetry shows it to have a hidden Type 1 nucleus
(Zakamska et al. 2005).}.
More details of the sample selection can be found in Appendix~A.   
Figure~\ref{fig:f24} shows the 24 $\mu$m flux distribution and spectroscopic status of the members of our sample.
Figure \ref{fig:f24_z_type1} shows the redshift distribution of the AGNs; virtually all 
of them are far behind the clusters. They are also far from the cluster 
centers  (typically by 1\arcmin~ or more), so they are not significantly magnified. 
We exclude any AGNs in the clusters according to the source redshift.

\subsubsection{Results\label{results_type1}}

 Of the 205 24 $\mu$m-selected Type-1
  AGNs we identified, 107 are securely detected by {\it Herschel}, 101  in at least two bands.
 For these sources, Table~\ref{table_fluxes_type1} presents coordinates, redshifts, and observed flux densities in the
  near-infrared ($J$ and $K$ bands from UKIRT and NEWFIRM), mid-infrared (3.4,
  4.6, and 12 $\mu$m from \wise; 24 $\mu$m from \spitzer), and
  far-infrared (100 $\mu$m, 160 $\mu$m, 250 $\mu$m, 350 $\mu$m, and 500 $\mu$m from
  \herschel). Similar information is provided in Table 4 for the sources not detected with {\it Herschel}.
  Below we discuss the basic properties of the {\it Herschel}-detected sources and
compare them with those not detected by \herschel\/ through a stacking analysis of the latter.
 In Section 3.5, we show that the SDSS colors of the \herschel\/-detected and \herschel\/-non-detected 
sample members do not differ significantly, showing that the dust emitting in the FIR 
is not producing significant extinction
along the line of sight toward the AGNs. This result is consistent
with the hypothesis that this dust is not associated with the active nuclei and simplifies the comparison 
of the {\it Herschel} detected and non-detected sources.  

\subsubsection{Completeness}

We ran simple simulations to test the completeness of the Type-1 AGN identification. 
The two plots in the upper panel of Figure
\ref{fig:k_histogram} show the expected apparent K-band and $r^{\prime}$ band (SDSS) magnitudes 
as a function of 24 $\mu$m flux density for Type-1 AGNs, represented by the template of \citet{elvis94}.
The K and $r^{\prime}$ band magnitudes are  functions of redshift and dust extinction. 
To simulate these effects, the Elvis et al. (1994) template was redshifted 
incrementally over the range $z=0$ to $z=3.6$, covering
the redshift range over which we identified AGNs.
We added extinction using a composite reddening law: 1.) the Galactic extinction
curve above 1 $\mu$m \citep{rieke85}; and 2.) a  SMC extinction curve below 
 1 $\mu$m \citep{gordon03}\footnote{Studies show that the reddening toward quasars is 
dominated by SMC-like dust at the quasar redshift (e.g. Richards et al. 2003; 
Hopkins, et al. 2004; also see the ``Gray" extinction curve in Czerny et al. 2004 and Gaskell et al. 2004)}; 
the value of $A_V$ ranged from  0 to 1.5.  
We rescaled the AGN template to make the 24 $\mu$m flux density always above 1 mJy.  
We used the same range of the flux level and $A_V$  
to generate the data points in the two plots in the upper panel of Figure
 \ref{fig:k_histogram};  in other words, the data points in these two plots
are from the same AGN populations. The simulation then lets us 
compare the incompleteness resulting from our detection limits for K-band and 
for optical spectroscopy.

Sources fainter than our K-band detection limit of 19 mag (Vega, 5-sigma) 
were not targeted in ACReS, but this K-band limit 
does not affect our AGN survey significantly.  As shown in the upper left of 
Figure \ref{fig:k_histogram},
all sources with 24 $\mu$m flux density above 1 mJy and  $A_V$  smaller than 1.5 mag have K-band magnitudes
brighter than 19 (Vega), and therefore were targeted.
The K-band limit would only exclude a few very red AGNs close to the 24 $\mu$m  1 mJy  flux density cutoff. 
The  $r^{\prime}$ band simulation shows that such very red AGNs would
drop below the limit for successful optical spectra. 
As can be seen from  the lower left of Figure
\ref{fig:k_histogram},  the K-band distribution of our 205 Type-1 AGNs 
declines steeply  from 17.5 to 19 mag (Vega), probably because the optical spectroscopy 
sets a tighter constraint for inclusion in our sample than does the K-band cutoff. 
As expected, reddening affects the $r^\prime$ band more than the K-band (see upper panels of Figure \ref{fig:k_histogram}). The lower right of Figure \ref{fig:k_histogram} compares the  
distribution of $r^{\prime}$ magnitude for targets that were put on Hectospec fibers
and targets where emission lines were detected (for sources with 24 $\mu$m flux density above 1 mJy).
The  redshift success rate starts declining from $r^{\prime} = 19.5$ mag (AB), 
and declines steeply for sources fainter than 20.5 mag (AB). 
We conclude that our spectroscopic survey is incomplete at the lowest 24 $\mu$m flux density levels 
for red sources with $A_V$ above 0.5 mag.

\begin{figure}[!hbt]
\centering
\includegraphics[width=1\textwidth]{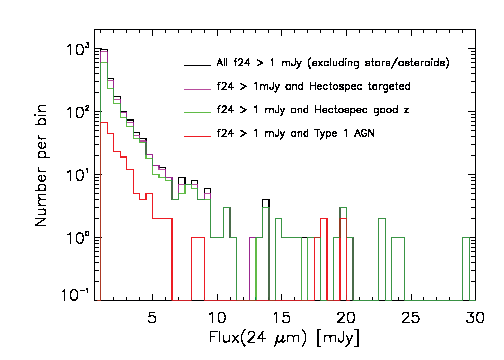}
\caption{24 $\mu$m flux density distributions of  sources 
 in our survey area. 
The black histogram shows the distribution of all sources with 24 $\mu$m flux density $> 1$ mJy 
excluding stars and asteroids.   
The magenta curve shows the distribution of sources targeted by Hectospec fibers (1729).
The green curve shows the distribution of sources with well determined redshifts from Hectospec spectroscopy (1209). 
The red histogram is Type-1 AGNs with emission line $\rm FWHM >1200 ~\rm km~s^{-1}$ (205).
}
\label{fig:f24}
\end{figure}

However, Figure  \ref{fig:k_histogram} implies that the completeness 
will increase rapidly as the 24 $\mu$m  threshold is raised above 1 mJy.
In confirmation, Figure \ref{fig:f24} shows that the fraction of Type-1 AGNs within 
the total sample targeted for spectra is roughly constant above 2 mJy, 
but is somewhat lower in the $1-2$ mJy bin. This drop toward 1 mJy is 
the behavior expected from incompleteness, but a part of the drop is also 
likely to be intrinsic, as shown by Brand et al. (2006). From Brand et al. 
(Figure 4, both data and Pearson models),
 we estimate that the intrinsic fraction at $1 - 1.5$ mJy should be about 70\% of the asymptotic  
value at larger flux densities. We then predict 100 AGNs in the $1 - 1.5$ mJy bin, 
where only 68 are detected, i.e., we are potentially missing about 30 AGNs. 
There is no evidence from the counts for any missing AGNs in the $1.5 - 2$ mJy or higher bins. 
We therefore estimate that the incompleteness in our sample is about 30/235 = 13\%, 
concentrated in the $1-1.5$ mJy range and largely due to incompleteness in the optical spectroscopy.  
Combining with the incompleteness of 5.4\% in the spectroscopy itself, the total of missing sources is
about 18\%.

In addition to the missing Type-1 AGNs, our sample will miss other 
types of active object. For example, from Dey et al. (2008), nearly half 
of the objects not targeted for spectroscopy because they did not have optical 
counterparts (see Appendix A) are likely to be dust-obscured galaxies. 
Because of their low accretion efficiency and low Eddington ratios, our sample
will also not include a significant number of jet-mode AGN \citep{yuan14}.
Our sample is therefore confined to traditional Type-1 AGN identified by 
optical spectroscopy, selected to a uniform level of mid-infrared flux density.

\subsection{Type 1 AGN Properties}

\subsubsection{Redshift Distribution \label{redshift}}

\begin{figure}[!hbt]
\centering
\includegraphics[width=0.7\textwidth, angle=90]{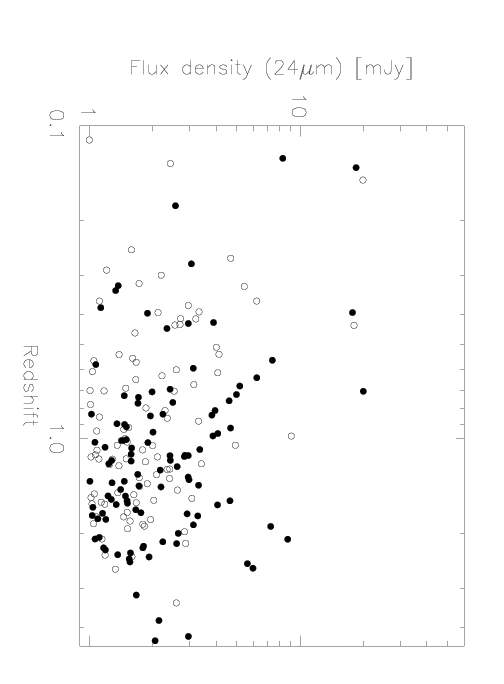}
\caption{24 $\mu$m flux density versus redshift for Type-1 AGNs with 24 $\mu$m flux density $> 1 $ mJy 
in our survey area. The filled circles show \herschel-detected Type-1 AGNs. 
The unfilled circles show \herschel-non-detected Type-1 AGNs. }
\label{fig:f24_z_type1}
\end{figure}

\begin{figure}[!hbt]
\centering
\includegraphics[width=0.37\textwidth,angle=90]{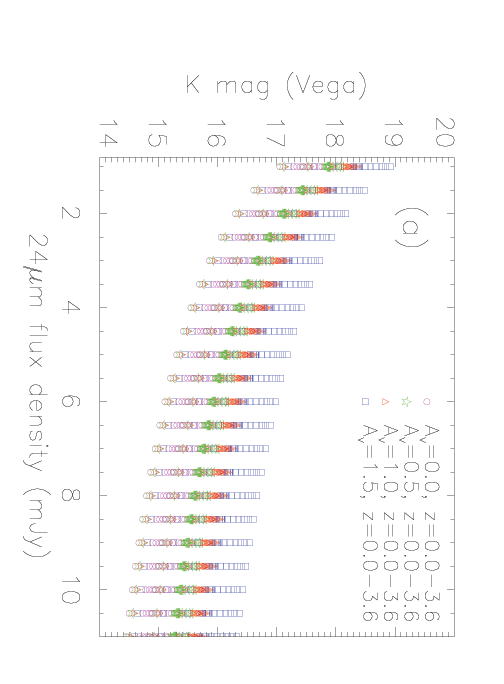}\includegraphics[width=0.37\textwidth,angle=90]{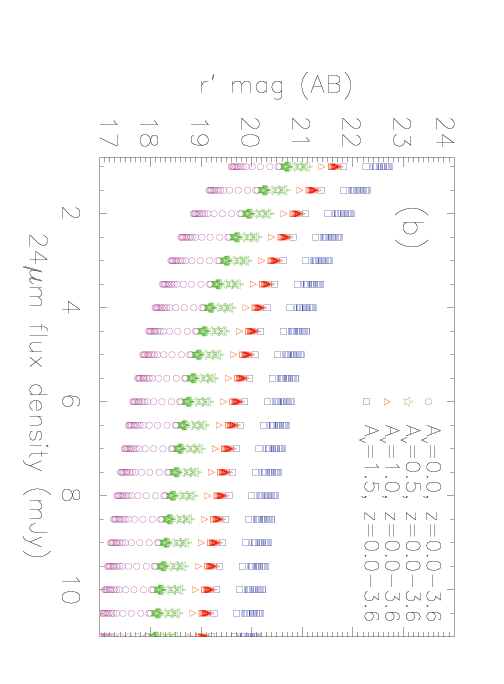}
\includegraphics[width=0.37\textwidth,angle=90]{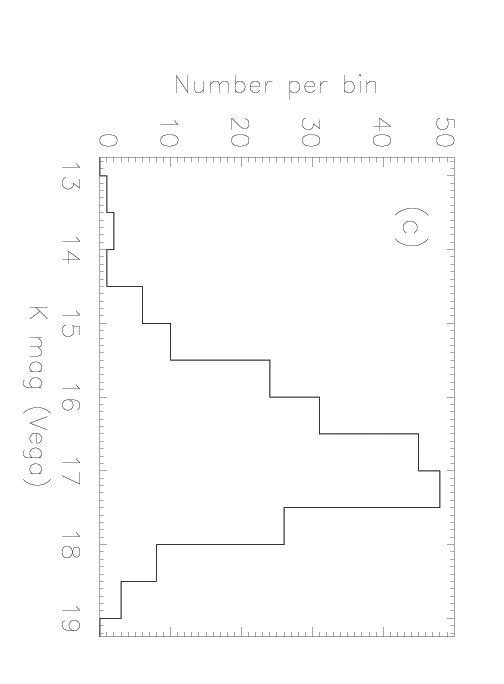}\includegraphics[width=0.37\textwidth, angle=90]{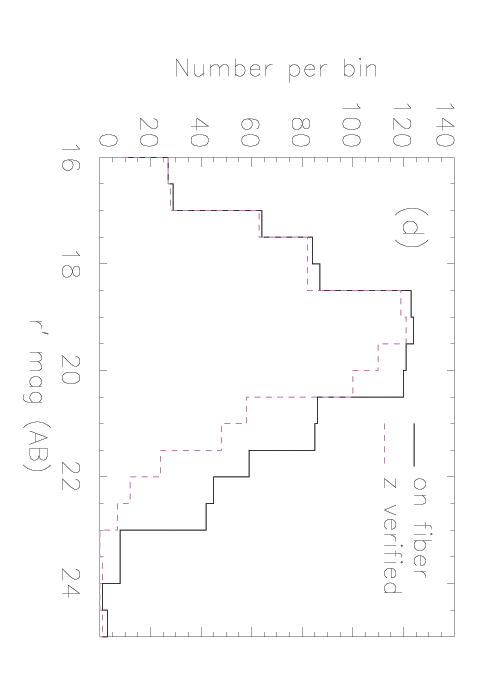}
\caption{(a): Simulation results showing the 
 expected apparent K-band  magnitude
as a function of 24 $\mu$m flux for Type-1 AGNs, 
as functions of  redshift ($z$ = 0--3.6) and dust extinction ($A_V$ = 0--1.5).
(b): Same as (a) but for  $r^{\prime}$  band.
(c): K-band magnitude (Vega) distribution of 205 Type-1 AGNs with 24 $\mu$m flux density $> 1 $ mJy  
in our survey area. 
(d): Distribution of $r^{\prime}$ band magnitudes for targets that were put on Hectospec fibers
and targets that are successfully identified with emission lines.}
\label{fig:k_histogram}
\end{figure}

Figure~\ref{fig:f24_z_type1} shows the redshift distribution of our Type-1
AGN sample\footnote{From SDSS spectroscopy, there are a number of sources
detected by {\it Herschel} at z $>$ 2; however, they are not included in our
study because at these redshifts the {\it Herschel} bands may be significantly influenced
by emission from the AGN.}.  
We omit four sources at $z>3$ (numbers 104 - 107 in Table 3) from further
analysis because of the very small number statistics in our sample
at these redshifts. The distributions for the \herschel-detected subsample and the
\herschel-non-detected one are very similar.  The two-sample
Kolmogorov-Smirnov test (K-S test) is consistent  at the 5\% level (P-value$= 0.053$) 
with the null hypothesis that these two subsamples are drawn
from the same distribution; that is, they are statistically indistinguishable.  
This situation is only possible if the luminosity of the FIR excess (which we will attribute to star formation)
grows with the increasing AGN luminosity that results from the flux limit of our AGN selection.

\subsubsection{Spectral Energy Distributions}\label{sed}

Three typical examples of the SEDs of these AGNs
are illustrated in Figure \ref{fig:sed_dcmp}.  
In $\nu f_{\nu}-\lambda$ units, some of the SEDs
look flat from the optical to the FIR, while some show peaks 
in the UV and optical, or in the FIR,  or both.  The FIR
peak has a Rayleigh-Jeans tail, declining steeply toward the mm-wave.
The SEDs of some sources at lower redshift show a peak near 1 $\mu$m
(rest-frame). 



By stacking signals over a large number of source positions, we can 
study the far-infrared properties of sources too
faint to detect with \herschel\/ individually.
We stacked the signals for  \herschel-non-detected Type-1 AGNs 
in three  redshift bins: 0.1--0.7, 0.7--1.2, and 1.2--1.9.
There are 24, 24, and 18 sources in these three  bins, respectively.
All the sources were well-detected from the UV to 24 $\mu$m. At these wavelengths, we took the average 
of the measured fluxes in each band for all of the sources in a redshift bin, after eliminating the 3-sigma outliers. 
For the five \herschel\/ bands, we registered PACS/SPIRE images with the 24 $\mu$m image 
by aligning the bright isolated point sources, and then isolated a small image centered on the
source position. 
We checked the images individually, and 
rejected those contaminated by close bright objects. 
For each redshift bin and each  \herschel\/ band, we then clipped the 3-sigma outliers for each aligned 
pixel over the full set of images. Since the \herschel\/ maps were roughly uniform in
exposure and noise, we could then take the straight average of all the remaining data.  Where there was a detection on the
stacked image, we did aperture photometry, including applying  
aperture corrections according to the parameters in Table \ref{tab:herschel_photo}.
The resulting  average SEDs in the three redshift bins (i.e., $z=$ 0.1--0.7, 0.7--1.2,
and 1.2--1.9) are shown in Figure \ref{fig:stack_sed_type1}, 
to be compared with the SEDs of the sources detected individually (e.g., Figure~\ref{fig:sed_dcmp}).  
The SEDs are similar, except the FIR peak is modestly weaker in the SEDs of the stacked signals.

\subsubsection{Nature of the Far-Infrared Excess}\label{nature}

We now explore the nature of the far infrared peak in the quasar SEDs. 
We display the [250/24 $\mu m$] flux ratio for the individual \herschel\/-detected objects 
as a function of redshift in
Figure~\ref{fig:fir_excess}, along with the upper limits 
for the \herschel-non-detected AGNs and the three average values for the stacked SEDs.  We also plot the flux 
ratios of a 
theoretical AGN template from Fritz et al. (2006) and a Type-1 AGN
template from \citet{elvis94}. The individual sources, upper limits, and stacked points all 
fall far above the model predictions. That is, AGN dust heating is unlikely to produce
adequate  FIR emission even if large (kpc-scale) tori are assumed  
\citep{fritz06, ballantyne06}. Therefore, the prominent far-infrared excess over 
the AGN templates indicates that a star formation component may 
contribute significantly to the FIR. Support for this conclusion is provided in
Section 3.4, where we show that the far infrared spectral energy distributions for most
of the sources are similar to those of normal star-forming galaxies of similar luminosity.

\subsection{The FIR Dust Temperature and Mass}

We estimate the temperature  of the FIR-emitting dust using a 
single-temperature grey-body model, of the form $B_\nu(T_\mathrm{d})[1-e^{-\tau_d}]$, 
where $B_\nu(T)$ is the Planck function and $\tau_d$ is the frequency-dependent
 dust optical depth. The dust is  optically thin in the FIR, 
and we have $1-e^{-\tau_d} \approx \tau(\nu) = \tau (\nu_0)(\nu/\nu_0)^\beta$.
Studies of local galaxies  \citep{hil83,dunne01,gordon10} 
show that a value of  $\beta  = 1.5$  is a good estimate of the emissivity index
for active star formation regions. 
We use the same criteria as  in \citet{Hwang10} 
to select sources with well-sampled SEDs around the peak of the FIR emission, 
i.e., there should be at least one flux measurement shortwards and longwards of the FIR peak, 
and the  FIR SED should be physical (convex, not concave).
There are 36 Type-1 AGNs in our sample that meet these conditions. 
The dust temperatures of these AGNs have large scatter from 22 K to 62 K. 
As shown in Figure~\ref{fig:dust_temp}, we compared the results with the 
luminosity-temperature relation for star-forming galaxies at $z=0.1-2$
derived from HerMES and PEP data in the COSMOS, GOODS-S and  GOODS-N fields  \citep{symeonidis13}.
The majority of the  AGNs with cold dust temperatures lie within the 1-sigma range of dust temperature
of star-forming galaxies, suggesting that the origin of their FIR emission 
is the same as for the cold dust in normal galaxies, that is star formation. 
Most such AGNs are at the lower luminosity end. The galaxies with
temperatures significantly above expectations for star-forming galaxies all 
have warm components (Section \ref{excess}) that make
determining the behavior of any cold dust ambiguous. 

For the galaxies where the FIR is dominated by the emission of cold dust, we can
estimate the required mass of interstellar material. In the optically thin limit, 

\begin{equation}
\label{eq:dustmass}
 M_\mathrm{dust} = \frac  {S_{\nu}{D_\mathrm{L}}^2 } {(1+z)\kappa(\nu_\mathrm{r}) B_{\nu_\mathrm{r}}(T_\mathrm{dust})}, 
\end{equation}

\noindent
where $S_{\nu}$ is the observed flux density at $\nu$ , 
$\nu_\mathrm{r}$ is the rest-frame frequency,  and the mass absorption coefficient, $\kappa(\nu_\mathrm{r}) = \kappa_0
(\nu_\mathrm{r}/\nu_0)^\beta$, is approximated by a power law.
Here we take $\kappa_0(125~ \mathrm{\mu m})=18.75~ \mathrm{cm^{2}g^{-1}}$ from 
\citet{hil83}.
The FIR-emitting dust mass ranges up to $9 \times 10^8 \rm ~ M_\odot $, 
with a median value of  $2 \times 10^8 \rm ~ M_\odot $, 
indicating huge reservoirs of gas in these systems. The large FIR luminosities 
for most of the other \herschel\/-detected systems also indicate large amounts of interstellar material.

\subsection{Type-1 SED Decomposition \label{seddcp_type1}}

We will show that the SEDs in  Figures \ref{fig:sed_dcmp} and \ref{fig:stack_sed_type1}
can be explained in terms of three dominant SED components; 1.) an averaged
Type-1 quasar continuum to supply the UV (the ``big blue bump" and to fill in the near-infrared; 
2.) the far infrared SED of a luminous star-forming galaxy; and 3.) the SED of a
moderately old stellar population, which peaks at wavelengths slightly longer than 1 $\mu$m.
These components are shown in the figures. In a few cases we need to add a warm far infrared component.
The overall simplicity of the SED (only three dominant components) underlies the
success of our fitting it to extract the underlying properties of the sources.
SED decomposition can determine quantitatively the relative contribution of the AGN and the
old stellar population in the NIR and of the AGN and star formation in the FIR.

\subsubsection{Decomposition Procedures\label{dcp}}

One of the uncertainties for the SED decomposition is 
the lack of a clean template of a naked Type-1 AGN SED. 
Numerical AGN models and  semi-analytic models provide candidate templates. 
The numerical models
assume a central point-like energy source 
with a broken power-law 
SED surrounded by a smooth or clumpy dust distribution, 
and then solve the radiative transfer equation (e.g. Fritz et al. 2006). 
Templates generated using this method must make assumptions about 
the dust distribution geometry and compositions. 
For semi-analytic methods (e.g. Mullaney  et al. 2011;  Sajina et al. 2012),
the SED is taken to be  a broken power-law based on 
physical assumptions for hot and  warm components, 
and  a modified blackbody beyond a given wavelength. 
Thus, both the numerical and semi-analytic methods are based on  assumptions 
and introduce a number of free parameters. 

To minimize the number of free parameters in our fits, 
we use an empirical AGN SED template \citep{elvis94} to determine the far-IR properties of our sources, and 
to estimate the relative contribution from the AGN and the host. 
This template may include a contribution from star formation and hence
be too bright in the far infrared. In Appendix C, we derive a correction to the
template that represents a bounding condition for maximal star formation. 
We begin with the published template and consider the implications of this 
second case in Section \ref{dcmp_result}. 

We model the observed SED  as 
a linear sum of a stellar component, a star formation component, and an AGN component.
We also assume the UV emission arises from the AGN rather than star formation.
The star formation and AGN are taken to be independent and not to affect each other.  
For the star formation component,
we use the 10 infrared galaxy templates from \citet{rieke09}
for luminosities of $10^{9.75}~ \rm  L_\odot$ to $10^{12} ~{\rm  L_\odot}$. 
This restricted range of IR luminosity has been shown 
to give appropriate FIR SEDs for galaxies at the redshifts in our study \citep{rujopakarn13}.
For the stellar component, we 
use 24 simple stellar population SEDs from Bruzual and Charlot (2003), 
 assuming a Salpeter IMF, Padova evolutionary tracks, and solar metallicity, 
at ages from 0.4 Myrs to 13 Gyrs.
Populations older than 1 Gyr have very similar SEDs in the NIR and beyond, with
any differences confined to shorter wavelengths (where the stellar output is
overwhelmed by that of the AGN). 
We also make sure the stellar component is not older than the age of the Universe. 
As necessary, we apply foreground extinction to the AGN template; we use the Galactic extinction
curve   above 1 $\mu$m \citep{rieke85}, and use a  SMC extinction curve below
 1 $\mu$m \citep{gordon03} (See Section 3.1.3 for more details about the
reddening model). 

We fit the above model to the data for rest wavelengths from $1216~\rm \AA$ to $1000$ $\mu$m (we do not
use the photometry short of 1216 \AA~ 
due to Lyman $\alpha$ forest absorption).
Specifically, we use photometry from $GALEX$, SDSS, J, K, WISE (3.4 $\mu$m, 4.6 $\mu$m, and
12 $\mu$m bands), 24 $\mu$m, \herschel\/ (100 $\mu$m, 160 $\mu$m, 250 $\mu$m, 350 $\mu$m, and 500 $\mu$m bands)  
if they are available. There are six degrees of freedom in the fitting: 
choosing the best fitting (1) infrared star formation and  (2) optical/near infrared stellar population templates 
from the libraries, (3) determining the extinction to the AGN, and normalizing (4) the Type-1 AGN, (5) star formation,
and (6) stellar population templates. We use Levenberg-Marquardt least-squares fitting to find the best 
solution among these degrees of freedom.

\subsubsection{SED Decomposition Results\label{dcmp_result}}

\begin{figure}[!hbt]
\centering
\includegraphics[width=0.4\textwidth, angle=90]{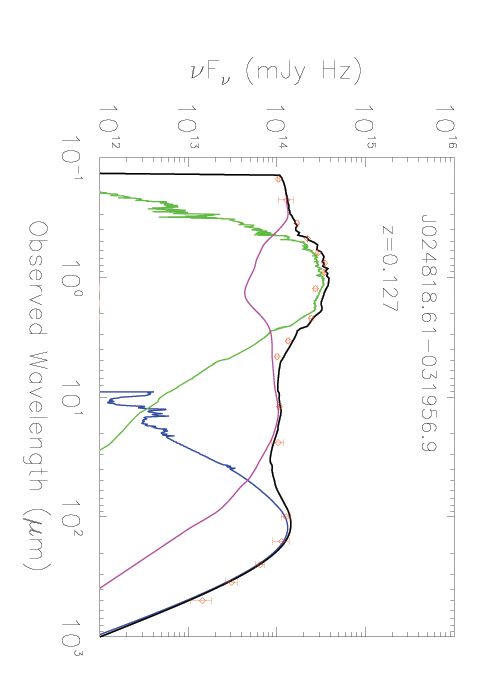}\\
\includegraphics[width=0.4\textwidth, angle=90]{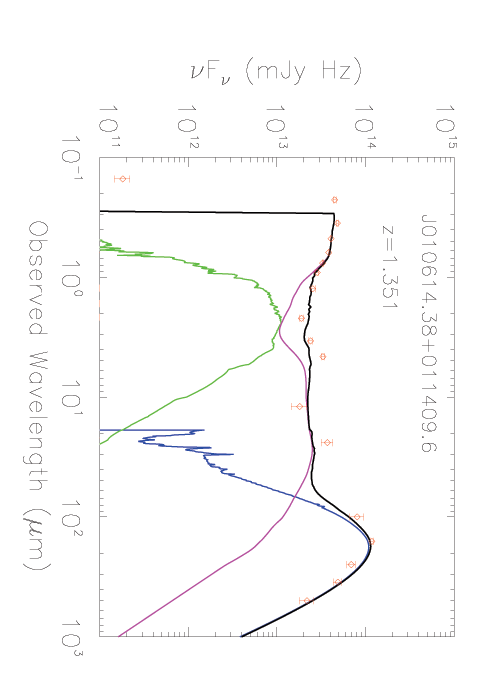}\\
\includegraphics[width=0.4\textwidth, angle=90]{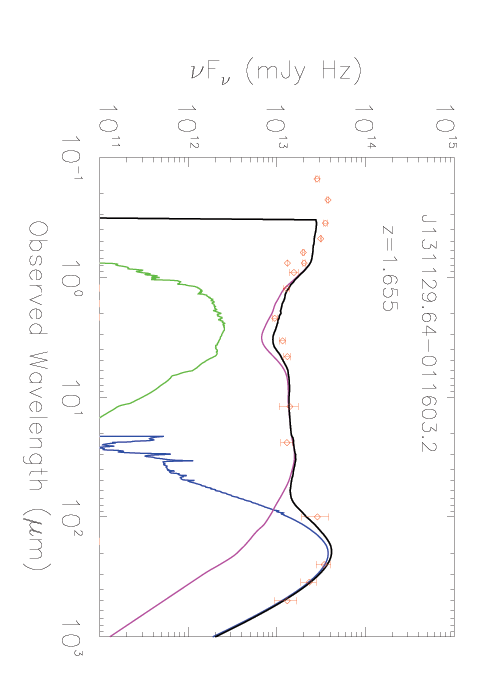}
\caption{Examples of SEDs and  decomposition results. 
 The diamond points are the average fluxes at FUV, NUV, and five SDSS bands 
(u$^\prime$, g$^\prime$, r$^\prime$, i$^\prime$, z$^\prime$),
J, K, and three WISE bands (3.4 $\mu$m, 4.6 $\mu$m, and 12 $\mu$m), 24 $\mu$m, and 
five \herschel\/ bands (100 $\mu$m, 160 $\mu$m, 250 $\mu$m, 350 $\mu$m, and 500 $\mu$m).
The solid lines show the SED decomposition results:
 the magenta line is the rescaled Type-1 AGN template \citep{elvis94}; 
the green line is the stellar photospheric component; the blue line is the best fitted star formation template.
The black line is the total of AGN, stellar, and star formation components.}
\label{fig:sed_dcmp}
\end{figure}

\begin{figure}[!hbt]
\centering
\includegraphics[width=0.4\textwidth, angle=90]{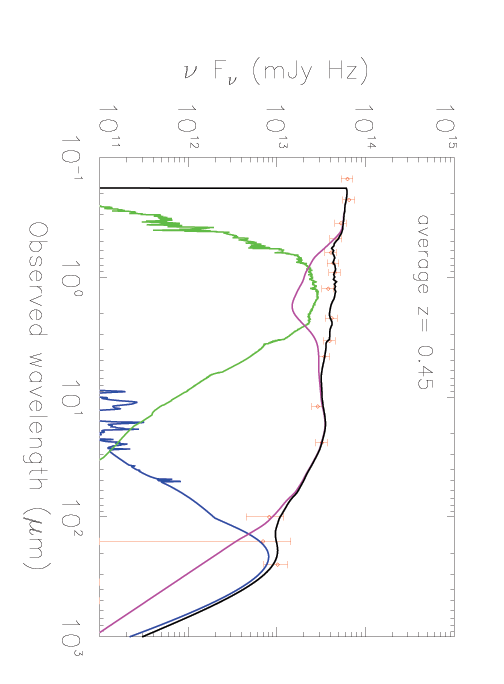}\\
\includegraphics[width=0.4\textwidth, angle=90]{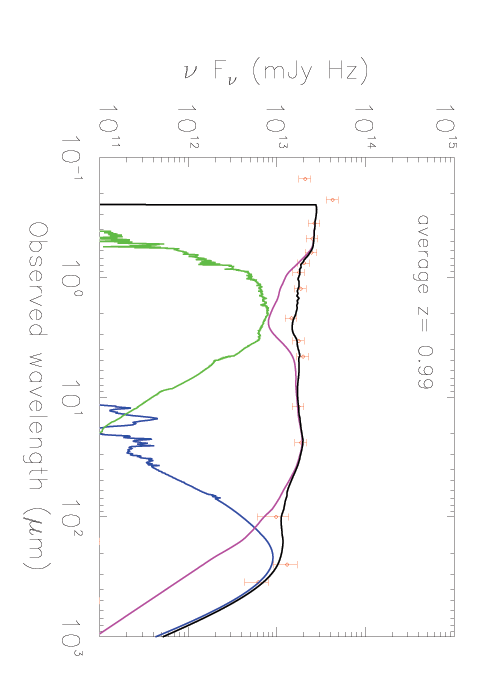}\\
\includegraphics[width=0.4\textwidth, angle=90]{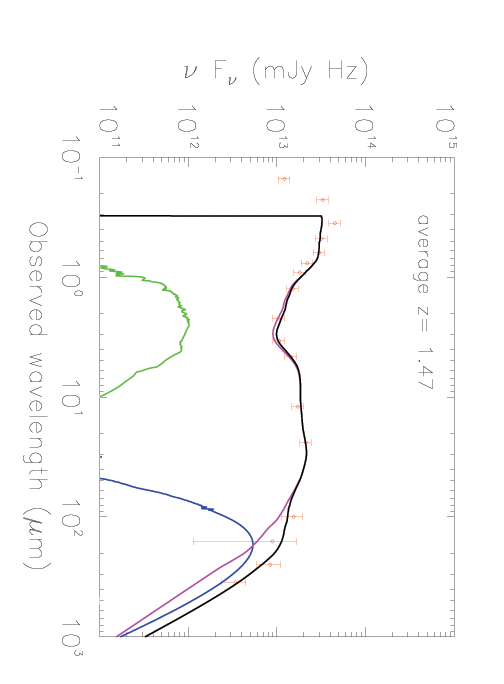}
\caption{The average SEDs for Type-1 AGNs with no formal FIR detections in our sample 
in three discrete redshift bins: $z=0.1-0.7$ (top, 24 sources), 
$0.7-1.2$ (middle, 24 sources), and $1.2-1.9$ (bottom, 18 sources). 
Symbols are the same as in Figure \ref{fig:sed_dcmp}.}
\label{fig:stack_sed_type1}
\end{figure}

\begin{figure}[!hbt]
\centering
\includegraphics[width=1.0\textwidth]{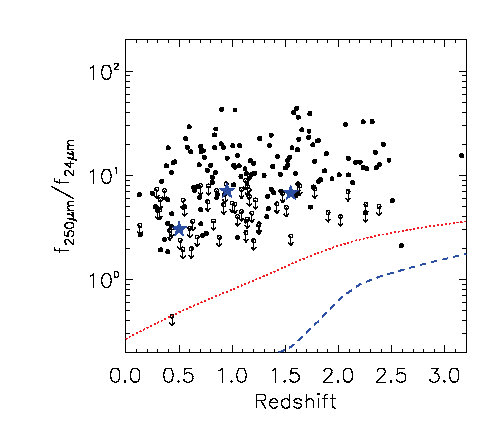}
\caption{
Observed-frame [250/24 $\rm \mu m$] flux ratio versus source redshift for the 
Type-1 AGNs in our sample. The downward pointing arrows indicate the sources 
not detected at  250 $\mu$m, based on $3\sigma$ upper limits. 
The five-pointed stars are the average values for \herschel\/ non-detected
Type-1 AGNs from our stacking analyses (see Section\, \ref{sed}).
The dotted (red) line is the [250/24 $\rm \mu m$] flux ratio of Type-1 AGN template
from \citet{elvis94}. The dashed (blue) line is the flux ratio of a typical Type-1 AGN template from \citet{fritz06}.}
\label{fig:fir_excess}
\end{figure}

\begin{figure}[!hbt]
\centering
\includegraphics[width=0.8\textwidth, angle=00]{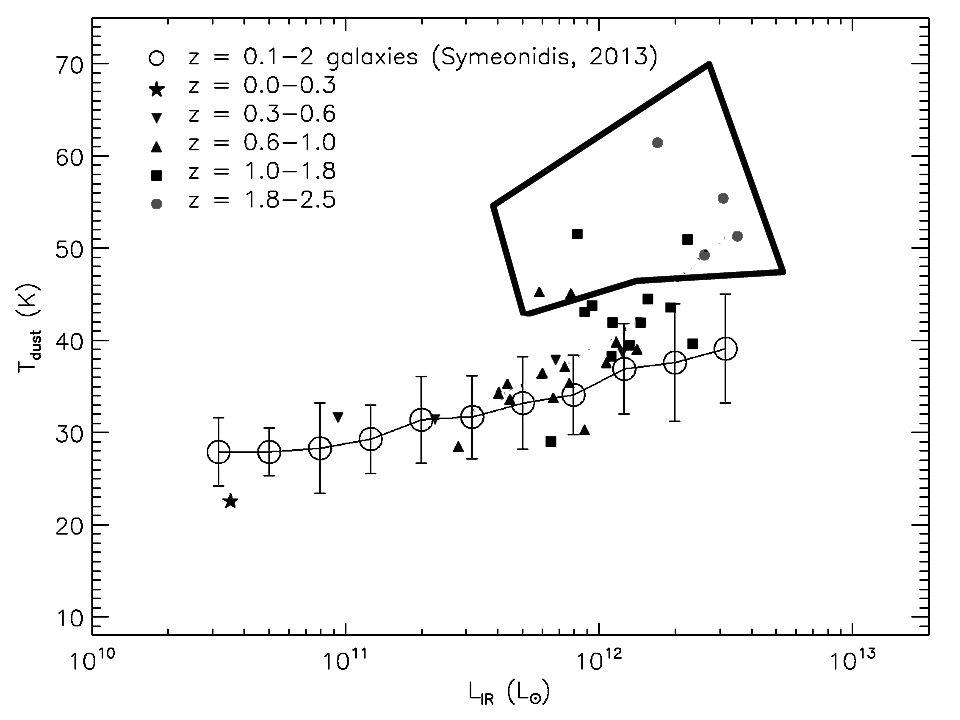}
\caption{The FIR dust temperature versus total infrared luminosity (rest-frame 8 $\mu$m-1000 $\mu$m).
The big circles (blue) show the L-T relation of star-forming galaxies (z=0.1-2) derived from Her-
MES and PEP data in COSMOS, GOODS-S and GOODS-N fields (Symeonidis et al., 2013). The
error bars show the 1-sigma scatter of the L-T relation. The blue irregular polygon encloses the
sources with strong warm infrared components. The remaining sources are compatible with the 
L-T relation, particularly if one allows for modest warm components in a few of them.
}
\label{fig:dust_temp}
\end{figure}

Figures \ref{fig:sed_dcmp} and  \ref{fig:stack_sed_type1} display examples of the SED decomposition results.  
In the rest-frame UV band and the MIR, 
the AGN  always dominates the emission for our sample. 
The UV and optical are dominated by thermal emission from an accretion disk, and a significant
portion of the NIR output is from hot dust warmed by the AGN. Many 
SEDs show a minimum at 1 $\mu$m (rest-frame), resulting from the upper 
temperature limit (sublimation temperature) for grains 
that can survive in the vicinity of an AGN. 
The stellar component contributes in the NIR, sometimes producing a NIR peak,
or making the SED flat near 1 $\mu$m.
The contribution from the stellar component is generally more significant at lower redshift. 
It fades quickly as the redshifts of the sources increase, as a result of our 24 $\mu$m selection
being dominated by the AGN and selecting increasingly luminous AGNs with increasing redshift.
A star formation component is needed for 95\% (102 out of 107) of the \herschel-detected sources,  but the
contribution of this component in the FIR varies substantially
from source to source. A star formation component is also required in all three redshift bins for the stacked SEDs 
of the sources not detected individually with {\it Herschel}. 

We have tested the necessity for star formation in the SED fits on 
the minority of sources where the FIR luminosity is
relatively small, namely 
quasars number 8, 24, 27, 38, 48, 49, 61, 64, 69, 71, 77, 90 (distributed over 
nearly the full redshift range of the sample, i.e., from z = 0.4 to z = 2.1). We
determined the minimum $\chi^2$ for fits to the measurements at rest-frame wavelengths $>$ 6 $\mu$m,
assuming 20\% minimum effective error (larger errors for low signal to noise measurements) 
for each photometric measurement and  
using just the Elvis template. We evaluated the quality of the fits based on the values of $\chi^2$ and the number of degrees of freedom for each galaxy, and then compared with the result with a star forming template added. The probability of the
fit being adequate with the Elvis template alone was $<$ 0.3\% in every case. With the star forming template added,
the probability that the fit was adequate was $>$ 15\% for seven (of twelve) cases, $\ge$ 1\% in three more, and was 
always much larger than the probability without the star forming template. The two cases with bad fits had far
infrared measurements that were incompatible with any smooth fit (i.e, the measurements indicated minima in the far
infrared, which is not a physically plausible behavior), suggesting that the issue is the data.
Thus, even for the individual systems where
we find relatively weak star formation, it is an essential part of the SED fits. This conclusion is
consistent with our finding in stacking the sources not detected individually that a star formation 
spectral component is present on average, although weaker than for the individual {\it Herschel}-detected
objects. 
   
A source of systematic error in the decompositions is the probable inclusion
of some far infrared emission due to star formation in the Elvis AGN template. We
have determined a bounding case (maximal level of star formation) for this effect as
described in Appendix C. Around 160 $\mu$m, this estimate attributes 75\%
of the template emission to star formation, so it is impossible to apply a substantially
larger correction. We have repeated the SED decomposition with this
star-formation-adjusted template and find for typical cases (where the FIR star formation component of
the decomposition is substantially stronger than the AGN template) that the upward correction in 
the star-forming luminosity is
only $\sim$ 10\%.    Larger corrections
apply for the twelve sources listed in the preceding paragraph with relatively 
weak star formation. Based on the
star-formation-adjusted Elvis template, the 
individual corrections to the estimated star forming luminosities for these
systems are 18, 9, 11, 17, 40, 11, 9, 17,  8, 48, 29, and 11\% respectively for galaxies 
8, 24, 27, 38, 48, 49, 61, 64, 69, 71, 77, and 90. 
For the stacked SEDs, the possible increases in the SFRs are
25, 10, and 46\% respectively, for 0.1 $<$ z $<$ 0.7, 0.7 $<$ z $<$ 1.2, and 1.2 $<$ z $<$ 1.9.
Applying these corrections would increase the necessity for star-forming templates in
fitting the {\it Herschel}-detected objects and would put the stacked results closer to the ones for the
{\it Herschel}-detected galaxies and emphasize that the non-detected galaxies are, on average,  
similar but modestly fainter in the FIR. 

There is another important conclusion indicated by  Figures \ref{fig:sed_dcmp} and \ref{fig:stack_sed_type1} and 
the SED decomposition. For all members of our sample, the 24 $\mu$m 
flux density is dominated by emission from the AGN, whereas the emission 
in the \herschel~ bands is dominated by star formation. Thus, 
the sample selection criteria are unaffected by the level of star 
formation in the host galaxies. The fact that the the {\it Herschel} 
detection rate does not fall significantly with increasing redshift indicates that both the AGN
and star forming luminosities in the sample increase with redshift at roughly similar rates, that is,
the host galaxy star formation must be roughly proportional to AGN luminosity (at least that at 24 $\mu$m).

\subsubsection{Warm Excess \label{excess}}

\begin{figure}[!hbt]
\centering
\includegraphics[width=0.5\textwidth, angle=90]{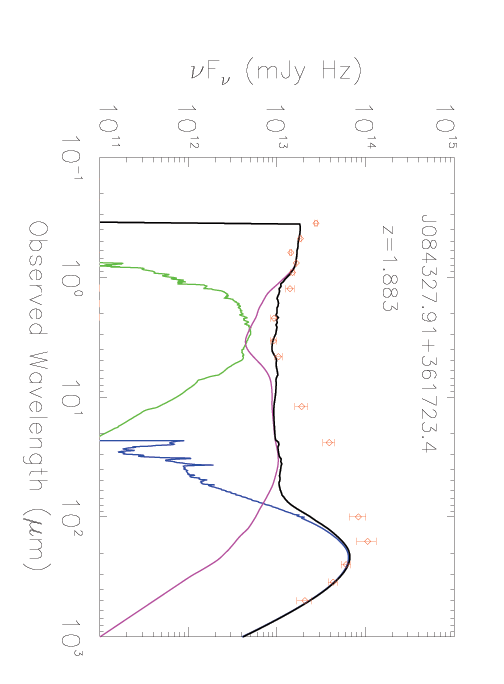}
\includegraphics[width=0.5\textwidth, angle=90]{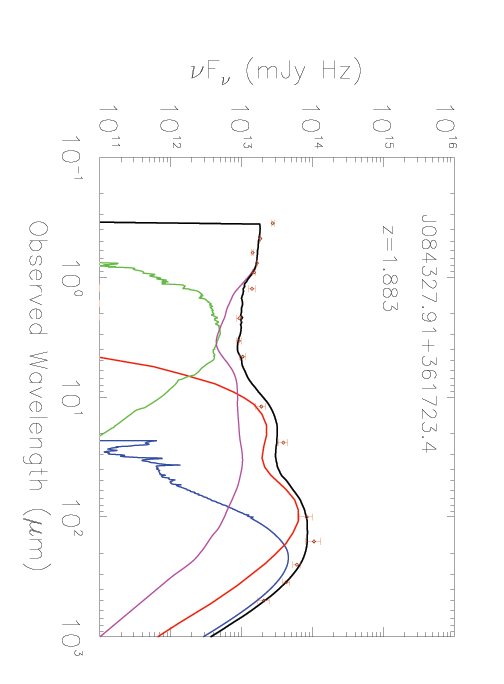}
\caption{SED decomposition results for an AGN  at $z=1.883$.
Top: Decomposition with 
stellar, AGN, and star forming galaxy templates only. 
The measurements from 10 $\mu$m to 200 $\mu$m (3 $\mu$m to 60 $\mu$m in the rest-frame) are high, indicating a warm excess 
from the MIR to FIR. 
Bottom: The same as the figure on the top. 
An additional warm component based on a model of a circumnuclear starburst 
has been added to improve the fit.
}
\label{fig:hot_excess}
\end{figure}

We found a strong excess above the SED decomposition 
result from  3  to 60  $\mu$m (rest-frame) for  some sources 
(see Figure \ref{fig:hot_excess} as an example); 
an additional warm component in addition to the star formation and 
AGN templates is needed to obtain 
a good fit. A similar excess 
is also found in some $z \sim 6$ quasar SEDs \citep{leipski13, leipski14}.
 A theoretical model of
a parsec-scale starburst disk (Thompson et al. 2005; Ballantyne 2008) 
predicts that a warm component heated by star formation would emit strongly in this wavelength range. 
To introduce a minimum of free parameters, we added a component with this specific spectrum 
(Ballantyne 2008, Figure 7) to the SED decomposition template library. 
There are of course alternative possible origins for this emission.
Figure  \ref{fig:hot_excess} shows the comparison in one example of the SED fits before and after adding the warm component.
The emission from the parsec-scale starburst disk reproduces the hot excess very well.
The total luminosity from the warm component for this source  accounts for 
 56\% of the total IR luminosity in this example. 

We judge the fits to require the warm component if the observed 12 $\mu$m, 24 $\mu$m and 100  $\mu$m fluxes
are about twice (or more) the fluxes from the SED decomposition 
only using AGN, stellar and star formation templates. The results are summarized in  Table \ref{table_luminosity_type1}. 
There are eight sources that require
a warm component to achieve a satisfactory fit, with a contribution to the total IR luminosity 
in the range of 30\% to 75\%. 

We also tested the influence of a warm component on the conclusions from all of the fits
where it did not appear to be required; the results are also in Table \ref{table_luminosity_type1}.
The column for $L_{\rm SF,IR}$ shows in parenthesis the fractional reduction in the luminosity of the
star-forming component if the warm component is added to the fit; for example, for source 1, the
fit is not improved with a warm component and there is no change in $L_{\rm SF,IR}$, whereas for source 2
the warm component improves the fit and reduces $L_{\rm SF,IR}$ by 10\%. In this latter case, the total
luminosity captured by the fit is also increased with the warm component, indicating that it accounts for
measurements that lie above the simpler fit.

For galaxies that are relatively faint in the far infrared, the introduction of a warm component can make
the optimum fit ambiguous. For example, of the 12 galaxies with relatively weak FIR discussed in the preceding
section, three (8, 24, and 64) could be fitted with a substantial warm component. Nonetheless, the purely
star forming FIR fits are also valid, all having probabilities $>$ 20\% of being satisfactory according to the
values of $\chi^2$. Given that the warm component seems to be prominent at high redshift and/or high luminosity, 
and that these galaxies have low far infrared luminosities and are at modest redshift, 
the star-forming template is the preferred fit. 

In summary, the decomposition inputs are surprisingly simple:
1) a fixed AGN SED  from \citet{elvis94} with adjustable foreground screen reddening; 
2) a far infrared SED appropriate for local LIRGs; 
3) a stellar population component $-$ although we allowed a broad of stellar population SEDs, 
the fits always converged on one appropriate for a relatively old population; 
 and 4) the warm IR component.

\subsection{Comparisons with Other Samples}

\subsubsection{Comparison of \herschel-detected  and Non-detected AGNs in the UV and Optical \label{redness}}

As discussed in Sec. \ref{redshift}, for the Type-1 AGNs there is  no obvious 
difference between the \herschel-detected
  and \herschel-non-detected populations in the 24 $\mu$m flux density distribution, 
nor in  the redshift distribution. We now expand that comparison to consider 
any differences in the UV and optical. The SDSS quasars are initially 
selected through a combination of optical colors and confirmed by optical 
spectroscopy \citep{richards02}. 
\citet{richards03} found that for them the relative color 
$\Delta (g^\prime -i^\prime)$ is a good indicator of quasar 
redness for redshifts between 0.3 and 2.2. 
The relative color is the difference between 
 the measured color of a given quasar and the median colors of quasars 
at the same redshift: a quasar with large $\Delta (g^\prime -i^\prime)$ 
could  either be intrinsically red or be reddened by dust.

\begin{figure}[!hbt]
\centering
\includegraphics[width=0.5\textwidth,angle=90]{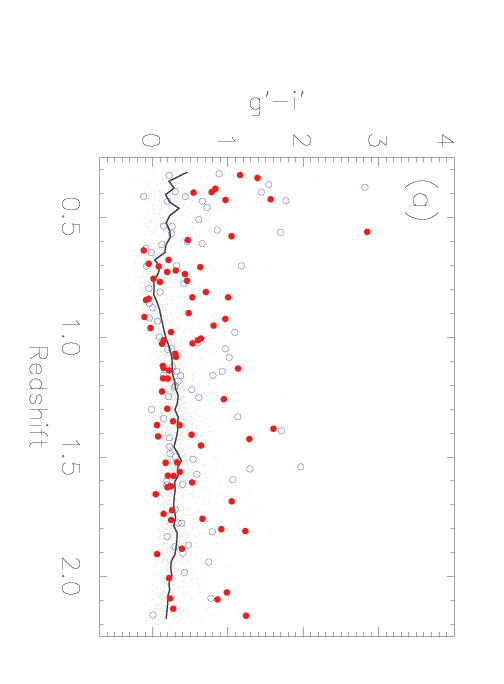}
\includegraphics[width=0.5\textwidth,angle=90]{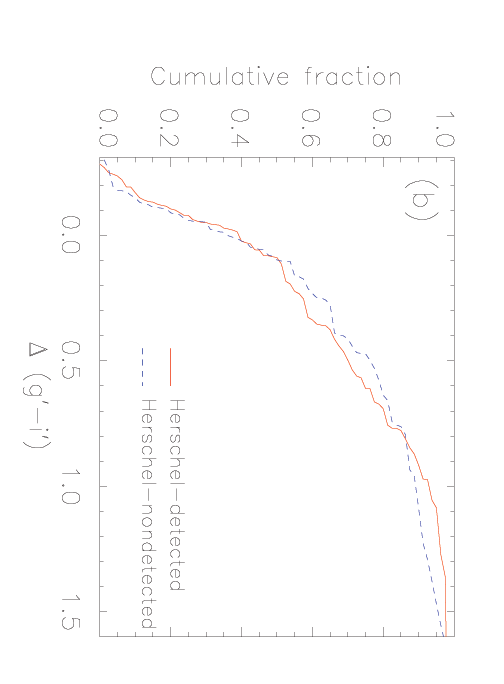}
\caption{ (a): The observed-frame  color  $(g^{\prime} -i^{\prime})$
of Type-1 AGNs in our sample as a function of redshift.
The filled circles (red) show \herschel-detected Type-1 AGNs; 
the unfilled circles (blue) show \herschel-non-detected Type-1 AGNs.
The small dots (grey) represent the SDSS optically selected Type-1 quasars from the 
SDSS Data Release 7 Quasar Catalog \citep{schneider10}.
The solid black line is the median value of the color of SDSS optically selected Type-1 quasars.
(b): The K-S test shows that the relative color distributions
of \herschel\/-detected (red solid) and -non-detected (blue dashed) 
AGNs are not statistically distinguishable (P-value=0.973).}
\label{fig:uvcolor_ks}
\end{figure}

There is SDSS coverage of 4.51 $\rm {deg}^2$ of our survey area. 
There are 84 SDSS optically selected Type-1 AGNs  in this area and 185 AGNs in our 24 $\mu$m selected sample.
61 AGNs are selected by both samples;  23 SDSS AGNs are not 
included in our sample due to their 24 $\mu$m flux densities being below 1 mJy.
The plot in the upper panel of Figure \ref{fig:uvcolor_ks} shows 
the colors of the 185 SDSS-and-24$\mu$m-detected quasars  
$(g^\prime -i^\prime)$  (corrected for Galactic extinction) compared with that of SDSS quasars in general.  
We determine the median colors of quasars at a given redshift (the solid line) 
using  data from the SDSS DR7 Quasar Catalog \citep{schneider10}. They represent 
a quasar population that is 
 optically-bright and not or only slightly affected by dust-reddening. 
The 24$\mu$m-detected quasars range from this line to being significantly ($\sim$ 1 magitude) redder; this
behavior is independent of far-IR properties.  
The K-S test shows that the relative color distributions 
of  \herschel\/-detected and -non-detected AGNs are not statistically distinguishable (P-value $=0.948$). 
In other words, \herschel-detected Type-1 AGNs are not significantly redder than
\herschel-non-detected ones in the optical.
This indicates that the dust responsible for the
  \herschel\/ detections is not producing any significant dust extinction
  along the line of sight toward the AGNs. This does not contradict our fits that
included reddening of the quasar template; it just indicates that the dust responsible
does not dominate the far-infrared emission.

\subsubsection{SDSS optically selected Type-1 AGNs\label{sdss}}

\begin{figure}[!hbt]
\centering
\includegraphics[width=0.5\textwidth, angle=90]{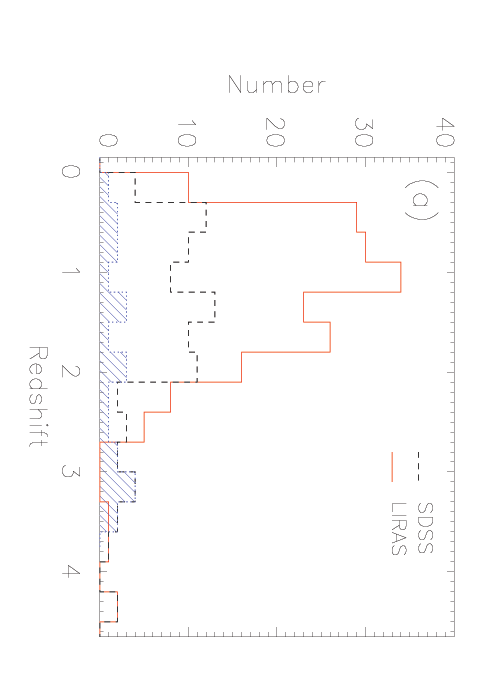}
\includegraphics[width=0.5\textwidth, angle=90]{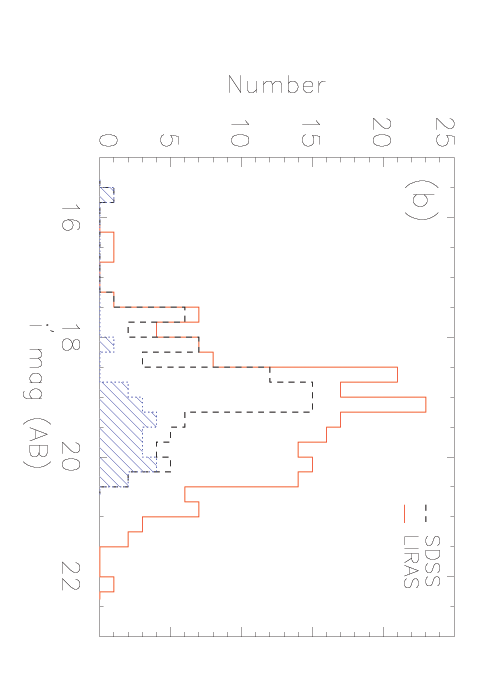}
\caption{ Comparison of  Type-1 AGNs in our sample and SDSS optically-selected Type-1 AGNs. 
(a): Redshift distribution. The red-solid histogram is Type-1 AGNs in our sample, while the black-dashed
histogram is the SDSS optically-selected sample.  The blue  hatched histogram is the SDSS AGNs that 
are not included in our sample due to their 24 $\mu$m flux density being below 1 mJy.
(b): $i^{\prime}$ magnitude distribution. The symbols are the same as  in the top panel. 
}
\label{fig:sdss_sample_comprison_hist}
\end{figure}

In Figure \ref{fig:sdss_sample_comprison_hist}  we compare 
the redshift and  $i^\prime$ band magnitude distributions for the SDSS and 24 $\mu$m samples. 
Basically the two samples select sources in the same redshift range. 
The 24 $\mu$m sample includes more  sources  faint in the $i^\prime$ band than the SDSS sample. 
The  23 SDSS AGNs that are  not included in the 24 $\mu$m sample
are evenly distributed over $z \sim 0-4$, and most of them lie at the faint end 
of the $i^\prime$ band magnitude distribution.
The majority of these  23 SDSS AGNs are detected at 24 $\mu$m, 
with flux densities in the range of $0.2-1.0$ mJy.
Therefore the bright 24 $\mu$m source selection has a large overlap with 
SDSS selection, but in addition finds many more (by a factor $\sim 2$)  Type-1 AGNs based on 
their optical spectra.

\begin{figure}[!hbt]
\centering
\includegraphics[width=0.5\textwidth, angle=90]{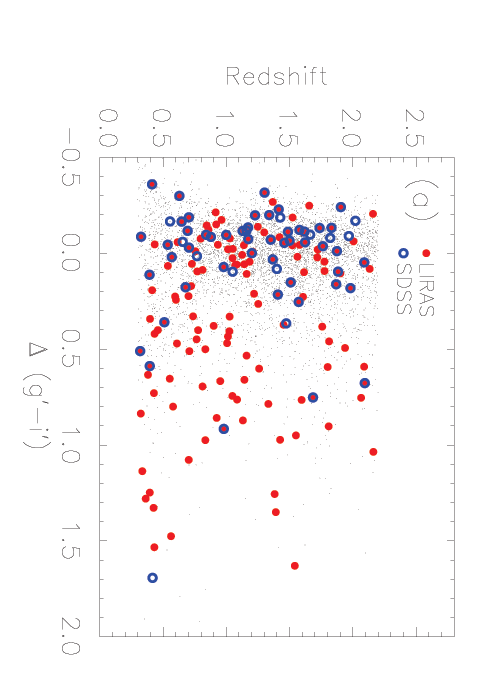}
\includegraphics[width=0.5\textwidth, angle=90]{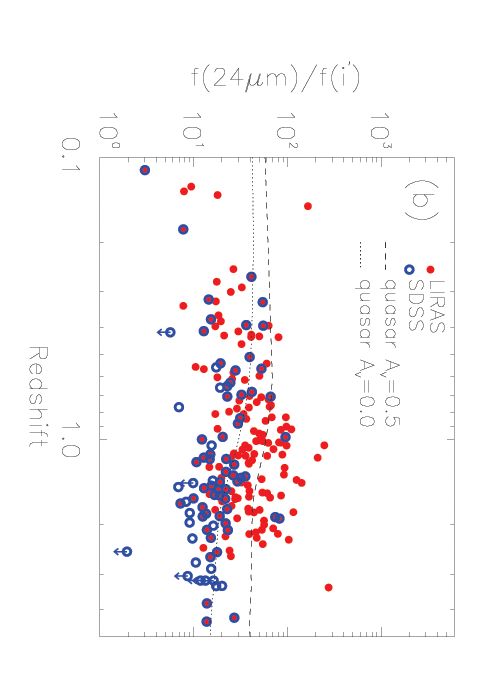}
\caption{ Comparison of  Type-1 AGNs in our sample and SDSS optically-selected Type-1 AGNs.
(a): The relative observed-frame color  $\Delta (g^{\prime} -i^{\prime})$ as a function of redshift.
The red filled circles  show Type-1 AGNs in our sample.
The blue circles  show the SDSS optically-selected AGN sample within the same area;
they are open if the source is not in our sample.
The small dots (grey) represent the SDSS optically selected Type-1 quasars from
SDSS Data Release 7 Quasar Catalog \citep{schneider10}.
(b): The [24 $\mu$m/$i^{\prime}$] flux ratio as a function of redshift.
Symbols are the same as in the upper panel. The dotted and dashed lines are 
the flux ratio calculated from Elvis's quasar template (1994) with reddening 
$A_{\rm V} =0$ and 0.5, respectively.
}
\label{fig:sdss_sample_comprison_color_ratio}
\end{figure}

In the upper panel of Figure \ref{fig:sdss_sample_comprison_color_ratio} 
we plot the relative color $\Delta (g^\prime -i^\prime)$ (see Sec. \ref{redness}) 
as a function of redshift for these two samples.
Most SDSS AGNs are scattered around the relative color 
$\Delta (g^\prime -i^\prime) = 0.0$.
This indicates that the SDSS AGNs in our survey fields
are typical of SDSS AGNs in general since the 
the median value of the color $g^\prime -i^\prime$ 
is calculated from the large SDSS Type-1 AGN sample. 
Our 24 $\mu$m selected sample includes additional red sources not identified by optical colors. 
At $z<1$, the red colors in our sample may partly arise from the contribution of the stellar component in 
the optical. 
Above $z\sim 1$, the 24 $\mu$m  selection picks up more luminous AGNs, 
and the SED is dominated by the AGN from the UV to the MIR; these red sources 
probably  either have strong  dust reddening or 
intrinsically red AGN continua. 

The lower panel of Figure \ref{fig:sdss_sample_comprison_color_ratio}
shows the observed-frame [24 $\mu$m/$i^\prime$] flux ratio 
as a function of redshift for these two samples.
The 23 SDSS AGNs that our 24 $\mu$m selection missed  
show small [24 $\mu$m/$i^\prime$] flux ratios due to their low 24 $\mu$m flux densities, 
and most of them are above $z\sim 1$. 
The contribution of the stellar component is also reflected in the  [24 $\mu$m/$i^\prime$] flux ratio
for sources at $z<1$ for both samples. 
For most of the AGNs at  $z<1$,  the [24 $\mu$m/$i^\prime$] ratios are lower 
than that predicted by a quasar template. 
At $z>1$, the SDSS-selected AGNs are more consistent with the quasar-template-predicted ratios 
with $A_{\rm V} =0$, while about half of the 24 $\mu$m-selected AGNs are more consistent 
with the template predicted ratios with $A_{\rm V} =0.5$. 
The redder color of [24 $\mu$m/$i^\prime$] for  24 $\mu$m-selected AGNs
may partly be due to dust-reddening or to intrinsically red continua in the optical, 
or partly due to the warm excess in the MIR enhancing the 24 $\mu$m flux density. 
In any case, these results demonstrate that 24 $\mu$m selection yields a 
substantial number of red quasars that are absent in purely optical selection.

\subsubsection{X-ray Selected AGNs\label{xray}}

We estimated the equivalent AGN X-ray flux to our 1 mJy selection threshold at 
24 $\mu$m using the  bolometric conversion from \citet{lusso12}, 
obtaining a flux of $10^{-14}~\rm erg~cm^{-2}~s^{-1}$  in 
the 0.5-2 keV band.
From the number of X-ray sources at a similar detection limit found by 
Cardamone et al. (2008), we expect a total of $\sim  260$ 
X-ray selected AGN above this flux level  in our total field. Our sample includes 205 infrared-selected sources, or
corrected for the $\sim$ 18\% incompleteness, about 240.  
However, from Cardamone et al. (2008), we expect the two samples to have different 
properties, despite the near-coincidence in numbers. 

The intrinsic X-ray luminosity in the [2-10] keV band 
for a typical AGN  at $z\sim 1$ in our sample is $2 \times 10^{44}~\rm erg~s^{-1}$,
  converted from a bolometric luminosity 
($\sim 5 \times 10^{45}~\rm erg~s^{-1}$ ) using the 
bolometric to  X-ray luminosity  correction in Figure 9 of \citet{lusso12}.
The failure to detect $\sim$ 80\% of the sample in the X-ray  
suggests that some members are moderately absorbed, consistent 
with their selection in the infrared. 
From these arguments, most of the AGNs in our sample are intrinsically more luminous in the X-ray, 
 compared with the X-ray selected, moderate-luminosity ($L_X = 10^{42}-10^{44} ~\rm erg~s^{-1}$) 
AGN sample in \citet{mullaney12}.
AGNs in both samples have comparable IR star formation luminosities and (specific) SFRs, 
and all reside in massive, main-sequence star-forming galaxies. 
However, in the optical and NIR, the SEDs of those X-ray selected, moderate-luminosity  
AGNs  are dominated by stellar emission (See the average SED in Figure 12 of  \citet{mullaney12}), 
while for our sample, emission from the AGNs is dominant in 
the optical and  NIR, except for some sources in the lowest redshift bins.
From Cardamone et al. (2008), the X-ray sample is expected to include 
nearly 50\% of sources where the near infrared is dominated by stellar 
emission, whereas the IR-selected sample is expected to be dominated 
by power-law sources in the near IR (See Donley et al. 2008). 
That is, a significantly higher fraction of the AGN luminosity emerges in the infrared for our infrared-selected 
sample than is the case for X-ray-selected samples.

\section{24 $\mu$m-Selected Type-2 AGNs \label{sec:type2}}

We now describe the Type-2 AGN identified from the same dataset. Many members of the Type-2 sample 
are at relatively low redshift and their AGNs tend to be of lower luminosity than the Type-1 objects. 
After finding the Type-2 objcts, we will identify a subsample that is directly comparable with the Type-1 AGNs in terms 
of redshift, black hole mass, and accretion rate. 

\subsection{Type-2 AGN Identification}

The sample of LIRAS
Type-2 AGNs is constructed based on the following selection criteria:
\begin{enumerate}

  \item  \spitzer/MIPS 24 $\mu$m flux densities above 1 mJy.

  \item  Optical spectra showing  narrow permitted emission lines 
	(full width at half maximum (FWHM) $< 1200$ \kms) with  high-ionization line ratios.

  \item If z $>$ 0.34,~ ${\mbox{\rm \nev}}$ $\lambda\lambda$3347,3427 detected or 
	FWHM(${\mbox{\rm \oiii}}$) $>$ 400 km s$^{-1}$.

\end{enumerate}

\noindent
Because the primary identification is based on emission line strengths rather than widths, 
well-calibrated spectra are required. To maintain consistency in
the classification, we only used Hectospec data; as a result, a few AGN identified in
SDSS that were not targeted with Hectospec may have been omitted from the Type-2 sample (see Tables 6 \& 7). 
We exclude any AGNs in the clusters according to the source
redshift.

A Hectospec fiber diameter subtends 1\arcsec.5 on the sky. 
At $z = 0.3$,  1\arcsec.5 subtends about 6.7 kpc and at $z = 0.6$, 10 kpc,
so the Hectospec fiber includes substantial light from the host galaxy.
The AGN emission lines can therefore be contaminated by stellar absorption lines from the galaxy.
Following Hao et al. (2005), we used the following procedures to subtract the host galaxy
contribution before measuring the AGN emission lines.
First, we select a sample  of  212 high S/N spectra of pure absorption-line galaxies 
from SDSS Data Release 7\footnote{http://www.sdss.org/dr7/}.
Second, we apply principal component analysis (PCA) 
to construct a library of galaxy absorption-line spectral templates.
Third, we fit a galaxy template, an A-type star template to account for the young stellar population in the host galaxy, 
and a power-law component proportional to $\lambda^{-\alpha}$ for the  nonthermal component from the AGN.
A $\chi^2$ minimizing algorithm was used  to determine the synthetic
stellar absorption spectrum.
Only after stellar and power-law continuum subtraction from all the spectra
do we measure the emission lines.

We fitted the following emission lines for each spectrum: 
  H$\alpha$, \nii$\lambda\lambda 6584,6548$, \hb, \oiii{$\lambda
5007$}. 
We rejected all objects with broad components (FWHM $> 1200$ \kms) in their emission lines 
(i.e., in \ha, \hb, or \mgii~2800). The minimum [OIII] line width 
criterion of 400 km s$^{-1}$ was based on the fitted width with no allowance for the 
spectral resolution. Given the resolution of R $\sim$ 1000, it corresponds 
to a threshold of about 270 km s$^{-1}$ for the intrinsic quasar line width. It should 
therefore not eliminate legitimate AGNs \citep{brotherton96} but protects against 
inclusion of chance anomalous star forming galaxies \citep[e.g.,][]{stanway2014}. 
We also rejected weak emission-line galaxies, i.e., 
the equivalent width of one of  \ha, \oiii, or \hb~
was required to be greater than 3 \AA.

There are several line flux ratio criteria to distinguish Type 2 AGNs from 
other narrow emission line objects (e.g., Kewley et al. 2001; Kauffmann et al. 2003).
Here we use the one from Kewley et al. (2001)
for objects at  $z<0.34$,

\begin{eqnarray}  
\log\left(\frac{\mbox{\rm \oiii} \lambda 5007}{ \Hbeta}\right) > \frac{0.61}{\log({\mbox{\rm \nii}/\Halpha})-0.47}+1.19.
\label{eq:kewleyline} 
\end{eqnarray}

\noindent
Since  $\mbox{\rm \nii}$ is redshifted out of Hectospec spectroscopic range at  $z>0.34$,
we use the following \citep{zakamska03} for objects at $0.34< z < 0.76$:

\begin{eqnarray} 
\log\left(\frac{{\mbox{\rm \oiii}}\lambda 5007}{{\Hbeta}}\right) > 0.3.
\label{eq:kewleyline4} 
\end{eqnarray}
 A few AGN at $z < 0.34$  are identified from the BPT diagram even if their \oiii ~ lines are
narrower than 400 km s$^{-1}$. Selection of Type 2 AGN with $z>0.76$ is not possible with 
our spectra since \oiii ~is redshifted out of the spectroscopic range.  

\begin{figure}[!hbt]
\centering
\includegraphics[width=0.7\textwidth,angle=0]{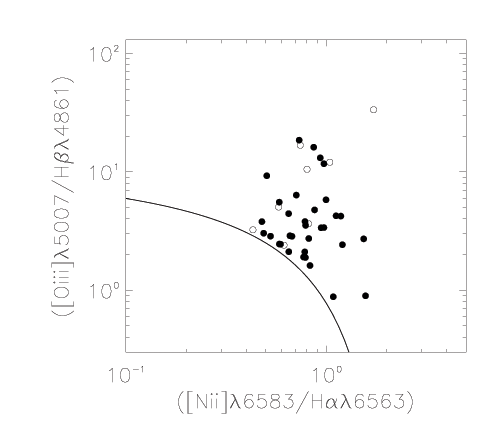} \\
\includegraphics[width=0.7\textwidth,angle=0]{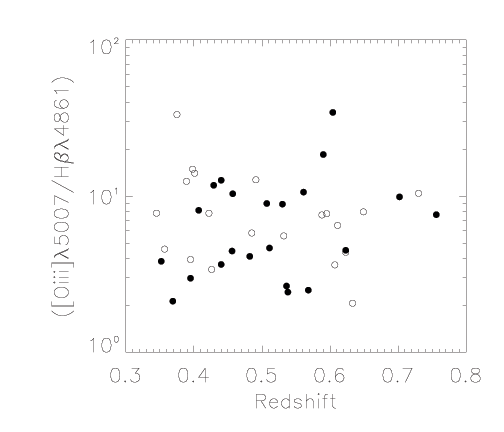}
\caption{Upper: Emission-line diagnostic diagram for sources at $z< 0.34 $ taken 
from \citet{kewley01} (Equation \ref{eq:kewleyline}). 
Filled circles are \herschel-detected Type 2 AGNs. Unfilled circles 
are \herschel-non-detected Type 2 AGNs.
Lower: The distribution of ${\mbox{\rm \oiii}}\lambda 5007/{{\Hbeta}}$ as a function of redshift 
for AGNs with z $>$ 0.34.}
\label{fig:bpt}
\end{figure}

The upper panel of Figure~\ref{fig:bpt} shows the  emission-line diagnostic diagram 
for Type-2 AGNs at $z<0.34$ selected in our sample using Equation \ref{eq:kewleyline}.  
The lower panel of Figure \ref{fig:bpt} shows the distribution of the \rm{[OIII]} to H$\beta$ line ratio for 
all selected Type-2 AGNs at $0.34<z<0.76$.

\subsubsection{Results}\label{result}

We identified a total of 85 24 $\mu$m-selected Type-2
  AGNs over the 3.6 deg$^{2}$  survey area; 55 are securely 
detected at least in two \herschel\/ bands, as listed in Table 6.
Figure \ref{fig:dcp_example} shows two typical SEDs for 
\herschel-detected AGNs.
The remaining 30 sources not detected with {\it Herschel} are listed in Table 7.
The redshifts and key derived parameters of the {\it Herschel}-detected objects can be found in Table 8.
We stacked the signals for  \herschel-non-detected Type-2 AGNs
in two redshift bins: 0.0-0.4, and 0.4-0.8.
There are 14 and 13 sources in these two  bins respectively, 
after rejecting those contaminated by close bright objects.
The stacked SEDs are shown in Figure \ref{fig:stack_sed}.


\subsection{Morphologies}

We visually examined the Suprime-Cam images (shown in Appendix A)  
to classify the {\it Herschel}-detected host galaxy morphologies (51 out of 55 sources have 
images of adequate quality). Although the images are not of sufficient resolution for a definitive determination 
of the morphologies of the entire Type-2 sample, they allow us to make plausible assignments for most members
(summarized in Table 8). Twelve of the 55 {\it Herschel}-detected galaxies either have no useful imaging
data (4) or are sufficiently compact that no further morphological information can be derived. Most
of the rest (38) are early-type spirals, lenticular galaxies, or elliptical galaxies. Only five are probable interacting systems, 
although a few of the ellipticals and lenticular galaxies also show hints of distortion and interaction. Therefore the majority of the AGNs reside in normal-appearing spheroidal and bulge-dominated galaxies.
This result is consistent with the results of 
Povi$\rm \acute{c}$ et al. (2012) for a sample of X-ray selected AGN at $z<2.0$ 
and with those reported by \citet{villforth14} for the CANDELS fields at $ z \sim$ 0.7.

\subsection{Type 2 SED Decomposition\label{seddcp}}

The Type-2 AGN sample does not extend to z $>$ 0.8 because the critical emission  lines move outside the range of our 
spectra; in fact, from the trend of detections with redshift, the sample becomes progressively less complete 
above z = 0.6. The sample includes many more galaxies at low redshift (z $<$ 0.3) than for the Type-1 sample. 
Many of these low redshift galaxies appear to be dominated by star formation at 24 $\mu$m, with AGNs both 
of relatively low luminosity (because of the low redshift) and, by themselves, not as bright as 1 mJy at 24 $\mu$m. 
To make these statements quantitative, we need to carry out the decomposition of the spectral energy distributions. 
Only then can we determine which sources can be compared directly with the members of the
Type-1 sample.

\subsubsection{Decomposition Approach}

For both the individual \herschel\/-detected sources and the stacked results, we 
 use SED decomposition to disentangle the AGN and star formation contribution 
in the FIR. Based on the arguments in Sections 3.2.3 and 3.3, we assume that star formation dominates
the signals in the {\it Herschel} bands.  Specifically, we model the observed SED  as
a  linear sum of a stellar component, a star formation component, and an AGN component.
We use Levenberg-Marquardt least-squares fitting to find the best
 stellar, star formation, and AGN templates and their normalizations.

The stellar population and FIR star-forming galaxy templates were identical to those used with the Type-1 sources. 
The \galex\ data show UV excess emission in the majority of the Type 2 AGNs that
cannot be produced by an old stellar population. 
This enhanced blue color is also reported in S$\rm \acute{a}$nchez et al. (2004) for 
early-type AGN hosts at $ 0.5<z <1.1$.
SDSS images of Seyfert 2 galaxies at $z<0.2$ show UV emission 
from young stars in the outer regions of the host galaxies \citep{kauffmann07}. Because the
stellar population SEDs did not allow for two distinct episodes of star formation widely separated in time, we added 
SEDs for a second population of very young UV-bright galaxies 
of age 0.1 Gyr to 1.0 Gyr to fit the UV emission. 

The strong extinction associated with Type-2 nuclei makes it impossible to use a single template 
to fit their SEDs. We employ the numerical AGN templates from \citet{fritz06},
which include cases with heavy absorption.
These  models assume a central point-like energy source
with a broken power-law SED surrounded by a smooth dust distribution,  
and then solve the radiative transfer equation.
Templates generated using this method depend on
the dust distribution geometry and composition, and the inclination of the torus toward the observer.
We put constraints on the AGN template library based on 
observations of the  Seyfert 2 galaxy  silicate 9.7 $\mu$m  absorption features, for which
it is found that $(F_f - F_c)/F_c > -0.85 $, where $F_f$ and $F_c$ 
are the observed flux density and underlying continuum flux density at the minimum of the 
silicate absorption feature, respectively  \citep{shi06}. 
Therefore, we do not use AGN templates with  silicate 9.7 $\mu$m  absorption 
$(F_f - F_c)/F_c < -0.85 $.
For comparison, \citet{nenkova08} have calculated models for clumpy dust
distributions; the comparison of smooth and clumpy dust models in \citet{feltre12} shows that,
although the two types of model give different outputs, for our purposes they
are equivalent. For example, the  \citet{nenkova08} models limit the silicate absorption
depth, which we did also by imposing an additional constraint on the \citet{fritz06} ones.

Samples of the deconvolutions are illustrated in Figures 13 and 14. Table 8 lists the
derived parameters for all 55 {\it Herschel}-detected sources. 

\subsubsection{Relative Roles of Star Formation and AGN at 24 $\mu$m}

\begin{figure}[!hbt]
\centering
\includegraphics[width=0.5\textwidth,angle=90]{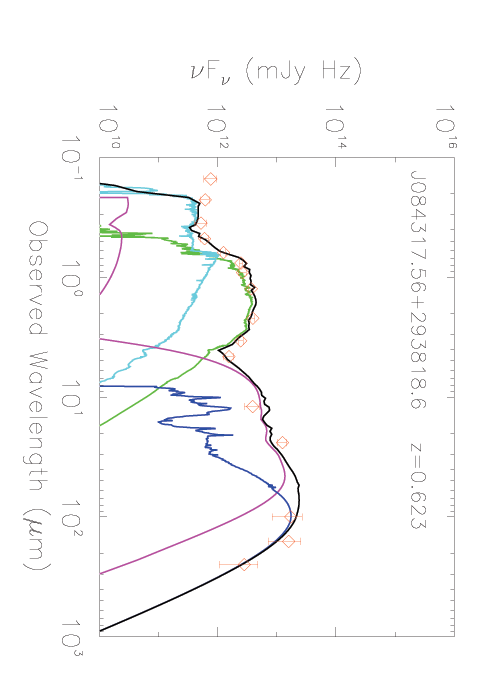}
\includegraphics[width=0.5\textwidth,angle=90]{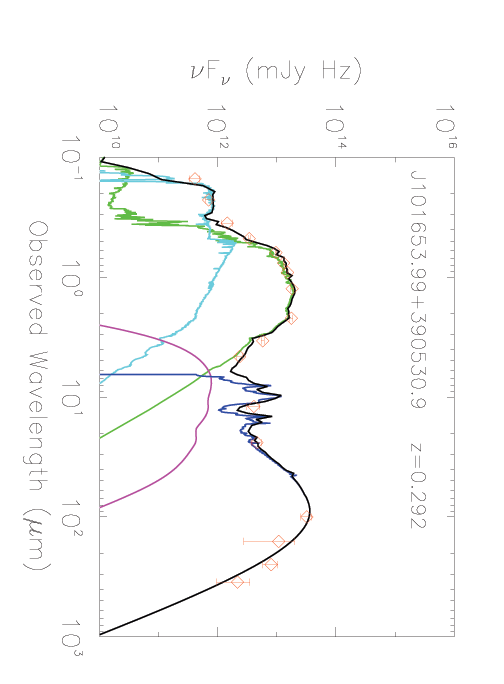}
\caption{Examples of SED decomposition fits. Upper: SED dominated at 24 $\mu$m  
by the AGN. Lower: SED dominated  at 24 $\mu$m by star formation.
The cyan, green, magenta, and blue solid lines represent  the best-fitting
young stellar component, old stellar component, AGN component,
and starburst component, respectively. The black solid line
represents the total of the best-fitting models.
}
\label{fig:dcp_example}
\end{figure}

All of the Type-1 AGNs are dominated by emission by the AGN at 24 $\mu$m 
(including the four at z $<$ 0.3)\footnote{In some cases, the warm component is dominant. The
origin of this component is not known, but it appears to be unique to AGN.}. We classify a Type-2 AGN as AGN-dominated if the flux 
arising from AGN component at 24 $\mu$m is larger than  that from the SF component; otherwise, it is defined as 24 $\mu$m SF-dominated. 
Figure \ref{fig:dcp_example}  shows examples of these two classifications. There are 17 AGN-dominated and
\herschel-detected Type-2 galaxies, about 1/3 of the \herschel-detected sample.  
There is an approximate divide in this behavior at z $\sim$ 0.3. Above this redshift, there are 50
Type-2 galaxies in our sample, of which 37 (74\%) are dominated (or tied) by emission by the AGN over that from
star formation at 24 $\mu$m (We include the {\it Herschel}-nondetected galaxies in this sample, since normalizing a star forming
template to the {\it Herschel} upper limits shows all of these to be AGN-dominated). However, for the 35
cases at z $<$ 0.3, only 13 (37\%) are AGN-dominated at 24 $\mu$m (including the {\it Herschel}-nondetected cases). 
In Section 3 we showed that selection at 24 $\mu$m yielded a large number of Type-1 AGNs; it appears 
that for z $>$ 0.3, selection at this wavelength is useful
to generate candidate lists that are relatively unbiased in terms of AGN type (see also \citet{mateos2013}). 

Because the 24 $\mu$m selection works relatively well at z $>$ 0.3 in finding the most luminous AGN,
we use it to compare the incidence of Type-1 and Type-2 sources. There are 50 Type-1 AGN with 0.3 $<$ z $<$ 0.8
in the 5.2 square degrees surveyed for them; normalizing by surveyed area, we expect 35 in the 3.6
square degrees surveyed for Type-2 sources. In fact, we have found 37 dominant Type-2 sources. That is,
the numbers of Type-1 and Type-2 quasars in this redshift range are similar. This result confirms the conclusion of
\citet{reyes2008}, but with the initial selection on a completely different basis than the extinction-corrected
[OIII] luminosity used in that work.

\subsection{Definition of the High Luminosity and Comparison Samples}

\subsubsection{Sample Definition}

We now derive a subsample of Type-2 objects suitable for comparison with the Type-1 AGN. The Type-1 sources are very luminous, with massive black holes (77/91 = 85\% have M$_{BH}$ $\ge 1 \times 10^8$ M$_\odot$) and accreting
at rates close to Eddington (66/70 = 94\% at $\ge$ 3\% of the Eddington rate)\footnote{The denominators for both of these
percentages are based on the number of objects with suitable measurements and do not represent the entire sample.}. 
To compare with them, we need to define a suitable sample of the 
Seyfert 2 AGN, namely those indicated to have M$_{BH}$ $\ge 1 \times 10^8$ M$_\odot$ and that 
are accreting at a minimum of 3\% of the Eddington rate, leading to a minimum bolometric AGN luminosity of
$10^{11}$ L$_\odot$. We describe these objects as the high luminosity sample (HLS) as indicated in Table 8.
The HLS consists of 17 objects, all but two at z $>$ 0.3. All but four of the 107 {\it Herschel}-detected
type 1 galaxies are also at z $>$ 0.3. \citet{yan13} show that the incidence of star-forming galaxies bright enough to
be within our 24 $\mu$m selection is 
low at z $>$ 0.3, simplifying the task of identifying AGN. 
Therefore, for the primary Comparison Sample with the Type-1 objects, we require z $>$ 0.3;
the 15 members of this sample are also flagged in Table 8.  

Not surprisingly
since both metrics emphasize high AGN luminosity,
13/17 of the AGN-dominated sources also belong to the HLS. By definition, the 15 sources in the 
Comparison Sample are all members
of the HLS, but the Comparison Sample also includes 14/17 of the AGN-dominated examples. 
Thus, the various methods for isolating the most luminous 
AGNs largely overlap. However, because its membership is not linked to the host SFR and its threshold AGN IR luminosity
is matched to that in the Type-1 sample, the Comparison Sample is best suited to complement the Type-1 sample.

\subsubsection{Possible Biases in the Comparison Sample}

The members of the Comparison Sample virtually all fall in the range where we identified the AGN by
the ratio of ${\mbox{\rm \oiii}}$ and H$\beta$ line fluxes. We now consider the reliability of this identification procedure. 
 Figure \ref{fig:l5007_lir_all} shows the correlation between the line luminosity, $L_{\mbox{\rm \oiii}}$, and the AGN 
total luminosity, the latter from our SED decomposition. 
The two luminosities correlate as $L_{\rm AGN} \propto  L^{0.74}_{\mbox{\rm \oiii}}$, even though it is generally
believed that the strength of the ${\mbox{\rm \oiii}}$ line should be proportional to AGN luminosity. 
At the higher end of the redshift range, sources have higher $L_{\mbox{\rm \oiii}}$ than 
pure proportionality predicts. Since $L_{\mbox{\rm \oiii}}$ traces ionizing photons that can be created by star
formation as well as AGNs, one possible reason for the $L_{\mbox{\rm \oiii}}$ excess at higher redshift is that the field of view of the fiber of the spectrograph includes significant amounts of ${\mbox{\rm \oiii}}$
emission from star formation in the host galaxy, as discussed further in Xu et al. (2015). 

The possibility, particularly at high redshifts, that our ${\mbox{\rm \oiii}}$ measurements are contaminated by 
the host galaxies could result in a bias against AGNs in host galaxies 
with very strong star formation, since they might be expected to have reduced ratios of ${\mbox{\rm \oiii}}$
to H$\beta$ and thus miss our selection criteria. However, we
believe this is not a problem for a number of reasons. First, we have searched for candidate 
contaminated systems at $z>0.34$ through the entire spectroscopic sample, by identifying galaxies
with ${\mbox{\rm \oiii}}$ FWHM $>$ 400 km s$^{-1}$, 1.5 $<$ ${\mbox{\rm \oiii}}$/H$\beta$ $<$ 2 (corresponding to 
0.176 $<$ log(${\mbox{\rm \oiii}}$/H$\beta$) $<$ 0.3, below the selection threshold in Equation \ref{eq:kewleyline4}), and
24 $\mu$m flux density $>$ 1 mJy.  Because all of our candidate galaxies have masses $> 
3 \times 10^{10} M_\odot$, this selection procedure would have
identified all Type-2 AGN in our stellar mass range that would satisfy the MeX criteria (see Figure 4c in \citep{juneau2011}). We found only one candidate. This low yield is consistent 
with our only identifying 5 out of our sample of Type-2 AGN above z = 0.3 with ${\mbox{\rm \oiii}}$/H$\beta$ between 2 and 3; it appears that our initial 24 $\mu$m selection generally yields AGN with relatively large ${\mbox{\rm \oiii}}$/H$\beta$.
This result suggests that there are very few candidates that might have missed identification as Type-2 AGN because
of contamination. The low yield with a relaxed  log(${\mbox{\rm \oiii}}$/H$\beta$) threshold also indicates
that the Type-2 sample is nearly complete to an AGN flux density of 1 mJy, at least for z $>$ 0.3. 

Second, consistent with this conclusion, if contamination were introducing significant
biases, one would expect that the \herschel\/-detected systems would have
a tendency to have low values of ${\mbox{\rm \oiii}}$/H$\beta$ because they have
relatively strong star formation, but the lower panel of Figure 12 does
not show a strong effect. 

Third, it appears that our AGN samples include all potential contaminated galaxies. 
We have determined that the ratio of ${\mbox{\rm \oiii}}$
to 24 $\mu$m flux density is roughly the same or slightly higher for AGN compared 
with star forming galaxies\footnote{For AGN, we used 
the sample from  \citet{diamond09} (nuclear 24 $\mu$m flux densities were provided by A. Diamond-Stanic,
private communication). We found an average value for the flux in [OIII] (in aW)
over the flux density at 24 $\mu$m (in Jy) of $513 \pm 80$ ($492 \pm 69$ for type 2 AGNs and $548 
\pm 118$ for type 1). For star forming galaxies, 
we utilized the MIPS 24 $\mu$m measurements and the "radial strip" line results for the SINGS sample
\citep{dale05,moustakas10} to find a ratio of $457 \pm 114$ and the integrated galaxy spectra
from \citet{moustakas06} with IRAS 25 $\mu$m data to obtain $426 \pm 15$. 
We have also used the "radial strip" spectra from \citet{moustakas10} for
galaxies with M$_V$ $<$ -20 to find an average value of ${\mbox{\rm \oiii}}$/H$\beta$ = 0.98 for luminous star-forming galaxies. 
Relaxing the luminosity threshold to M$_V$ $<$ -19 has little effect: the average is then 1.03, so the value is
not strongly sensitive to galaxy luminosity (and the accompanying range of relevant metallicity, based on the
luminosity-metallicity relation). \citet{caputi08} find a similar average ratio, while the work of \citet{moustakas06} yields
a value of 0.82, again in good agreement. }.
In addition, the value of ${\mbox{\rm \oiii}}$/H$\beta$ intrinsic to AGN is often significantly higher
than our adopted theshold of 2. As an example, for the sample compiled by \citet{lamassa10}, the median ratio is 9, 
while the sample of \citet{juneau2011} has a typical ratio of 4. 
Taken together these results show that any host galaxy containing an AGN with an intrinsic flux 
density $\ge$ 1mJy at 24 $\mu$m (i.e., above the luminosity threshold for our AGN samples)
plus star formation sufficiently vigorous to 
contaminate the ${\mbox{\rm \oiii}}$/H$\beta$ ratio enough to cause it to fall below our threshold 
would have a total signal at 24 $\mu$m well above 1 mJy. However, above 2 mJy, the AGN in our samples
count for all of the detections, leaving no room for a population of luminous AGN in very luminous 
star-forming galaxies.

Figure \ref{fig:l5007_lir_all} is not the first finding of a departure from the expected 1:1 relation between ${\mbox{\rm \oiii}}$ and bolometric AGN luminosity in the direction of an increasing [OIII] lluminosity for more luminous AGN. 
\citet{lamassa10} found similar behavior relative to 12 $\mu$m luminosities; 
Shao et al. (2013) also saw this behavior when comparing the [OIII] and 22 $\mu$m luminosities 
for a large sample of AGN from the Sloan Digital Sky Survey; and Hainline et al. (2013) report a similar departure from a 1:1 relation using 8 $\mu$m luminosity as an indicator of AGN luminosity. However, Shao et al. (2013) show a 1:1 relation between [OIII] and 4.6 $\mu$m luminosity. 
Taken together these results imply discrepancies in measuring the AGN luminosities from 
different single-color infrared bands. It is therefore of interest that we find the effect based on the bolometric AGN luminosity, 
rather than an estimate of the luminosity based on a single spectral band.

\section{Intrinsic Properties \label{estimation_type1}}

With the SED decompositions, along with other derived properties of the sources (e.g., line widths), we now
derive the basic physical parameters of the sources in our sample.  

\subsection{IR Luminosities and Star Formation Rates\label{sfr}}
The SED decomposition disentangles the contributions from different source components. 
The infrared luminosity from the star formation component ($L_{\rm SF,IR} $) is integrated
over the rest-frame $8-1000$ $\mu$m range of the best-fit star formation template. 
The infrared luminosity from the AGN component ($L_{\rm AGN,IR}$) 
is integrated over the same range of the rescaled AGN template\footnote{The calculated 
luminosities are uncertain for a number of reasons, such as: 1.) the underlying assumption that the
emission by the central engine is isotropic, despite its complex geometry and optical depth (e.g., \citet{koratkar99}; 
2.) the contamination of the AGN template in the FIR by star formation (see Appendix C); and 3.)
the inclusion of {\it both} the optical-UV-X-ray and the infrared components of the SED \citep{marconi04}. 
It is difficult to make quantitative estimates of these effects, but other than the first, they appear to
be modest (i.e., $<$ a factor of two for the second two together (see \citet{marconi04} and Appendix C).
}. By integrating the full Elvis template, we set the total AGN luminosity of the Type-1 objects to be 5.28 times their infrared luminosities. The Type-2 AGN bolometric luminosities are
taken to be the rescaled total intrinsic luminosity of the best-fit AGN template for each source\footnote{Uncertainties 
for them include: 1.) the differences between clumpy
and smooth models (Feltre et al. 2012); 2.) in our fitting, the torus opening angle is poorly constrained; 3.) variability;
and 4.) the underlying assumption that the emission by the central engine is isotropic, despite its complex geometry
and optical depth (e.g., Koratkar \& Blaes (1999))}. 
The total infrared luminosity ($L_{\rm IR}$) is the sum of $L_{\rm SF,IR}$
and $L_{\rm AGN,IR}$. 
The star formation fraction ($F_{\rm SF}$) is defined 
as $L_{\rm SF,IR}/L_{\rm IR}$. 
Of the  \herschel-detected Type-1 sources, $ 21 \%$ have $F_{\rm SF} > 75\%$,
$ 47 \%$ have  $50\% <F_{\rm SF} < 75\%$, 
and $ 32 \%$ have $F_{\rm SF} < 50\%$. The corresponding values for
the Type-2 Comparison Sample are 60 $\pm$ 20\%, 27 $\pm$ 13\%, and 13 $\pm$ 10\%, similar within the poor
statistical weights of the latter (particularly allowing for the lower typical redshifts of the
Type-2 objects). 
The  star formation fractions
of the stacked SEDs of \herschel\/-non-detected Type-1 AGN
(in three discrete redshift bins: $z=$0.1--0.7, 
0.7--1.2, and 1.2--1.9) are all below $40\%$.

The star formation rates are calculated from $L_{\rm SF,IR}$
using the relation in \citet{kennicutt98}, adjusted
for a ``diet" Salpeter initial mass function (IMF) \citep{bell03} from
the original Salpeter IMF, i.e.,

\begin{equation} \frac{{\rm SFR}}{{\rm  \rm M_\odot~yr^{-1}}} =
  1.2\times10^{-10}\left(\frac{L_{\rm SF,IR}}{{\rm \rm L_\odot}}\right).
\label{eq:sfr_lfir_eq}
\end{equation}

\noindent
The adopted IMF reduces the proportion of low mass stars to resemble, for example, the Kroupa
IMF, and puts the SFRs on the same scale as our mass estimates in the following section. 
$L_{\rm SF, IR}$, ranges from $\sim 10^{10}$ to $3\times 10^{12}  ~\rm L_\odot$ for 
the {\it Herschel}-detected galaxies; the average value for the stacked SEDs 
of the non-detected galaxies are several times lower. Nonetheless, star formation activity must be common 
even for the AGN hosts not individually detected by \herschel. However, as for the 
local sample of PG quasars (Shi et al. 2014), it 
is possible that there are a number of quiescent galaxies among those we stacked, and
therefore that elevated star formation is not ubiquitous.

\subsection{Virial Black Hole Masses and Eddington Ratios\label{bhmass}}

Type-1 AGN black hole masses,  $M_{\bullet}$, have been measured directly by 
reverberation mapping (Blandford \& McKee 1982;
Peterson 1993; Kaspi 2000),  
but it takes years to obtain results using this technique.
However, reverberation mapping has also provided empirical scaling relations
allowing us to estimate black hole virial masses efficiently from
the quasar continuum luminosity and  broad emission
line widths, e.g., H$\beta$ (4861 \AA), Mg {\sc ii} (2800 \AA) , and C {\sc iv} (1549 \AA). 
We used the moderate resolution ($\sim 6~\AA$, corresponding to 300-400 km s$^{-1}$) 
Hectospec spectra to determine the FWHM 
 of the broad emission lines. 
We followed the procedures in \citet{vestergaard01} (for Mg {\sc ii}), 
and \citet{peterson04} (for H$\beta$ and  and C{\sc iv}) to fit these lines
and measure the FWHM of the broad component. 
We took the  mass-scaling relationship from \citet{vestergaard06} (for 
 H$\beta$ and C{\sc iv}) and from \citet{vestergaard09} (for Mg {\sc ii})
 to  estimate black hole masses. 
In Appendix \ref{black_hole_mass}, we  list the three mass-scaling relationships 
we used, and show three examples of fitting results for 
\mgii, \civ, and \Hbeta, respectively\footnote{If both  \Hbeta~ and \mgii~ are available, we adopt \Hbeta; 
and if both \mgii~ and \civ~ are available, we adopt \mgii. }. 
The measured FWHMs and estimated BH masses  
are listed in Table \ref{table_luminosity_type1}. 

The Eddington Luminosity from a black hole with mass $M_{\bullet}$ powered by spherical accretion is 
\begin{equation}
L_{E} = \left(\frac{4\pi Gcm_p }{\sigma_e}\right)M_{\bullet}. 
\label{eq:qso_eddington0}
\end{equation}
We obtained the AGN total luminosity from the SED decomposition, and calculated the 
ratio of AGN luminosity to Eddington luminosity. 
Of the Type-1 AGN, 94\% emit at $\gtrapprox$3\% of the Eddington Limit. The distribution of the bolometric luminosity as a fraction of the Eddington limit 
is consistent with that of the SDSS quasars \citep{mclure04}.

Assuming the local stellar mass ($M_*$) and black hole mass ($M_\bullet$) correlation 
(i.e., $ M_* \approx 700 M_\bullet$; e.g., Bennert et al. 2011; Cisternas et al. 2011; Scott etal. 2013), we calculate the black hole masses, 
Eddington luminosities, and Eddington ratios of  the Type-2 AGNs, with results
shown in Figure \ref{fig:stellar_blackhole_eddington}.
The 24 $\mu$m SF-dominated Type-2 AGNs have slightly lower ratios than the AGN-dominated ones at 
all redshifts. The 24 $\mu$m AGN-dominated Type-2 galaxies emit close to 10\% of the Eddington rate (14/17, or 82\% emit
at $\gtrapprox$3\% of the Eddington rate); for z $>$ 0.3, their behavior is similar to that of the Type-1 AGNs. 
Therefore, as expected (e.g., by the unified model), the behavior of the nuclei of 
 the 24 $\mu$m AGN-dominated Type-2 galaxies is consistent with that of the Type-1 sample. 
The lower Eddington ratios in the 
star-formation-dominated galaxies are expected, given that they have not been selected strictly on
AGN luminosity.

\subsection{Stellar Masses of AGN Host Galaxies\label{stellarmass}}

\subsubsection{Stellar masses from SEDs}

Based on the SED decomposition, we 
can estimate the host galaxy stellar masses.
Because the details of the stellar spectrum are difficult to disentangle from the AGN emission, 
we base the mass estimate on the NIR stellar luminosity, which has been shown 
to be an accurate approach \citep[e.g.,][]{mcgaugh2014}. 
We use the relation between stellar mass, $M_*$, and K-band luminosity, $L_k$, for
local field galaxies \citep{bell03}: 
($\rm log_{10} (M_*/L_k) = -0.42 + 0.033 ~log_{10} (M_{\rm c}h^2/\rm  M_\odot$), where 
$M_{\rm c}h^2$ is 10.63 averaged over all galaxy types
 (10.61 for early-type galaxies and 10.48 for late-type galaxies). The masses 
assume a "diet" IMF, defined by \citet{bell03}. We need to be sure 
that our mass estimates are on a consistent scale with other approaches. This
would be straightforward if the host galaxies were normal early-types, but many of them
have anomalously blue colors \citep[e.g., ][]{floyd2013}. We therefore compare with a wide
range of masses that include galaxies with a range of colors. We find that the 
masses are consistent with those using SDSS KCORRECT \citep{blanton07}\footnote{Also see
http://howdy.physics.nyu.edu/index.php/Kcorrect. The SDSS KCORRECT stellar mass is 
based on the Bruzual-Charlot stellar evolution synthesis and makes use of the multi-band SDSS photometry.}
for local galaxies (private communication, Krystal Tyler; See Figure \ref{fig:stellarmass_comparison}). Although
photometrically determined masses can be subject to significant systematic errors, the 
agreement on average between our approach and the masses derived from full photometry puts our masses on
the same scale as, e.g., those of \citet{elbaz11} and allows direct comparison of the host galaxy behavior
in our sample with the field galaxy behavior described in that paper.
The stellar component in the Type-2 galaxies is more accessible to our fitting than for the Type-1 cases, 
and we could attempt more detailed models. However, for consistency in 
comparing the samples, we use the same approach.
 We obtain the estimates of stellar mass in Table 8.
Our AGNs reside in very massive galaxies
 with stellar masses around $10^{11} \rm ~ M_\odot$.

As a stellar population ages, its luminosity declines as its more massive stars die; 
 i.e., a fixed K-band luminosity corresponds to smaller stellar mass at higher redshift. 
This passive evolution must be accounted for in estimating the masses of the stellar
populations in high redshift galaxies (e.g., Drory et al. 2003, 2004; van der Wel et al. 2006; 
also see van Dokkum \& Franx, 2001). 
We assume the AGN host galaxies evolve passively and follow van der Wel et al. (2006)
to correct for this systematic evolution of the host stellar luminosity 
(i.e., $\Delta~ln~(M_*/L_K) = (-1.18 \pm 0.10)z$). 
This correction can be applied to galaxies from the local epoch to $z\sim 1.2$, 
where the correction factor is equal to 4.1.
For $z\sim 1.2-2$, we keep the value of the correction factor at 4.1. 

Since the rest-frame J-band is near both the peak of the stellar SED and a minimum of the AGN SED,  
we use this band to quantify the stellar component output. 
Our fits constrain the J-band flux from the stars well for most low-z sources, where
the AGNs are of relatively low luminosity and do not dominate in the rest NIR. 
At higher redshifts, we can usually only obtain 
upper limits for the stellar fluxes. 
We ran a simulation to test to what level we can trust the stellar flux from 
the SED decomposition. 
First, we renormalized the stellar and AGN SED templates to the desired 
flux ratio in the rest-frame J-band. 
Second, we applied dust extinction selected randomly over the range $A_{\rm V}$ = $0 - 1.0$ 
to the AGN template and then added the two
templates.  Third, we convolved the bandpass transmission curves
with  the combined templates to simulate the photometry that we 
used for the SED decomposition. We added random noise to the simulated photometry 
in all bands, assuming a standard deviation of 20\% in consideration of the photometry errors, AGN variability, and that the 
data in different bands were probably taken in different years. 
Fourth, we ran the SED decomposition procedures on this simulated photometry 
and calculated the recovered flux ratio of the AGN and stellar components
in the rest-frame J-band.
The input flux ratio was set to six discrete values:  $\rm flux_{AGN,J}/flux_{Stellar,J} =0.5, 1, 2, 3, 4, 5$, and 
the calculation was repeated 10,000 times for each value.
We compare the input and recovered flux ratios in Figure \ref{fig:simulation_ratio_agnstar}. 
If the stellar flux  is twice the AGN flux in J band, 99\% of the sources can be recovered accurately. 
If the stellar flux  is equal to the AGN flux in J band, there is a larger scatter of the 
recovered flux ratio, and on average, the stellar mass is overestimated by about 20\%.
However, $> 95\%$ of the sources are recovered within a factor of two. 
If  the stellar flux  is  below the AGN flux in J band, the errors in the stellar flux 
are large. Based on this result, if the rest-frame 
J-band flux of the stellar component is equal to or above that 
of the AGN component, we can compute a valid stellar mass.  
If the J-band flux of the stellar component is smaller than that of the AGN component, 
we use the  J-band flux of the AGN component to assign an upper limit to the
mass of the stellar component. 

The $J-K$ colors of early-type galaxy stellar populations are very similar, so the rest-frame K-band 
luminosity can be taken to be 0.85 times the J-band luminosity. 
Therefore we use 0.85 times the stellar component J-band flux to compute stellar masses, or
of the AGN component to estimate stellar mass upper limits.  
We then obtain the estimates of AGN host stellar mass as tabulated in Table 5.

\subsubsection{Indirect determination of stellar masses}

We now estimate stellar masses from the black hole-stellar bulge relation. These estimates
allow us to

\begin{enumerate}

\item extend the study of AGN host galaxies to a significant number at z $>$ 1

\item test the passive evolution assumed to correct our mass estimates from observed near infrared fluxes

\item investigate the possible bias toward massive host galaxies because requiring them to be 
sufficiently bright in the NIR to outshine the AGNs for photometric mass estimation will favor ones
 with massive hosts, at least at high redshifts

\end{enumerate}
\noindent
To lay the foundation for indirect mass estimates, we: 1.) examine the possible
extent of evolution of $M_{\bullet}/M_*$ over the relevant redshift range, 0 $<$ z $\le$ 1.8; and 2.) 
calibrate the masses derived from $M_{\bullet}$ against those from near infrared luminosity 
obtained in the preceding section. These
two steps let us determine the maximum plausible deviations of the derived stellar masses from a nominal
$``$best estimate." 

The great majority of luminous AGN are in galaxies with early-type morphologies \citep[e.g.][]{mcleod1995, floyd2004}. 
For such galaxies, the local value of $M_{\bullet}/M_*$ is well determined for galaxies with  $M_{\bullet}$ $> 3 \times 10^7$ M$_\odot$ \citep{kormendy2013}. The majority of our
AGN samples with z $\ge$ 0.3 have $M_{\bullet}$ above this threshold, within the range where $M_{\bullet}/M_*$ is
well behaved. 

Most investigators agree that, within the errors, there is little evolution in the $M_{\bullet}/M_*$ ratio from z = 1 to z = 0 \citep{peng06a, shen08, somerville09, cisternas11, zhang12, salviander13, salviander14}. A small number of studies suggest some evolution in this range but are inconclusive regarding its significance \citep{woo08, canalizo12}. We will assume that the local ratio holds up to z $\sim$1.2. For z between 1 and 2, the indications range from very little evolution \citep{peng06b, jahnke09, somerville09, sarria10, schulze14} to evolution by a factor up to about four (at z = 2) \citep{peng06a,trakhtenbrot10, merloni10, decarli10, bennert11}. Particularly at z $>$ 1, there are selection effects that bias the apparent evolution upward, correction for which reduces it significantly, to a factor of two or less \citep{lauer07, shen10, schulze11, portinari12}\footnote{The factor of two is also consistent with the conclusion that half of the stellar mass observed today has formed since z = 1.3 \citep{madau2014}, 
assuming that any black hole growth over this period is negligible.}.  However, these same selection effects, e.g. the bias due to luminosity selection toward relatively massive black holes \citep{lauer07}, may apply to our use of the $M_{\bullet}/M_*$ ratio
to estimate stellar masses\footnote{See \citet{matsuoka14} for an example of about a two times bias for galaxies of similar mass to ours.}, so for our sample, we will consider the possible extreme value of  $M_{\bullet}/M_*$ at z = 1.8 to be 4 times the local value.

To calibrate stellar masses from $M_{\bullet}/M_*$  against masses from near infrared photometry, we assume the $M_{\bullet}/M_*$ relation  has no evolution from  
the local value $M_* \approx 700  M_{\bullet}$ up 
to $z \sim 1.2$ (e.g., Bennert et al. 2011; Cisternas et al. 2011; Scott et al. 2013).
At $0<z<1.2$, there are 28 AGNs in our sample that have both $M_{\bullet}$ 
(derived from broad line width; See Section \ref{bhmass})
 and $M_*$ (estimated via K band luminosity). 
In Figure \ref{fig:stellar_mass_more}\footnote{One extreme outlier has been omitted; 
such outliers are also seen in other samples \citep{kormendy2013}}, 
we compare the K-band derived $M_*$'s and those from $M_{\bullet}/M_*$.
The large scatter is not surprising given the scatter 
in the $\sigma$-M relation \citep{kormendy2013} and in the mass determinations 
from photometry (e.g., \citep{shapley01, savaglio05, kannappan07}), plus 
the additional uncertainties indicated by our simulation of the deconvolution uncertainties in near infrared fluxes. 
The K-band derived $M_*$  is on average two times higher than the bulge stellar masses predicted 
by the local $M_{\bullet}/M_*$ ratio.
This offset is redshift independent, which indicates our correction for passive K-band 
evolution is roughly correct. 

The offset could arise if the galaxies have substantial disks, or if 
our photometric or BH masses have small systematic errors. However, there is also a 
 selection bias toward relatively massive galaxies that have sufficiently bright NIR 
stellar fluxes to outshine their AGN and allow mass estimates from their SEDs.  Approximating this 
bias by assuming a normal intrinsic distribution with all of the cases with 
mass estimates from SED fits coming from the upper side indicates an offset by 
a factor of 1.7, in satisfactory agreement with what we find. Thus, the indirect stellar mass
estimates serve the important function of removing this source of bias from our sample. The scatter 
is 0.38 dex rms\footnote{This value is plausible given the expected errors in single-epoch black hole 
mass estimates found by \citet{vestergaard06}, corrected for scatter in 
the reverberation mapping masses \citep{onken04}, or the single epoch error limits estimated by \citet{denney09}}.  
We can estimate the intrinsic scatter 
in our near-infrared-based mass estimates of about 0.15 dex from  Figure \ref{fig:stellarmass_comparison} 
corrected for the scatter in the masses with full photometric fits \citep{santini14}. If anything, 0.15 dex 
may be a low estimate \citep{courteau14}.  We need to add
the scatter due to having to measure the near infrared fluxes from deconvolution of 
SEDs with significant contributions from the AGNs. The resulting total scatter is expected to be 
at least 0.2 dex. Subtracting this value quadratically from 0.38 dex, we estimate that
the masses determined from $M_{\bullet}/M_*$ will have an intrinsic rms
scatter of 0.32 dex.  Given the uncertainties 
we estimate the possible errors as 0.3 dex toward low values (from the rms scatter) but 
0.6 dex toward high values above z = 1.1  (from possible evolution and/or selection effects), relative to  
the nominal values assuming  the local $M_{\bullet}/M_*$ ratio. All the AGNs with stellar masses from photometry reside in very massive galaxies 
 with stellar mass around $1 - 4 \times 10^{11} ~\rm  M_\odot$. The masses 
estimated indirectly are also generally within this range. 


\section{Summary\label{conclusion_type1}}

We studied the properties of a sample of 24 $\mu$m-selected, 
spectroscopically-identified AGN and their host galaxies, using  a 
multi-wavelength dataset from  $Chandra$, $GALEX$, SDSS, UKIRT, WISE,
$Spitzer$/MIPS, and \herschel.
Typical luminosities for these AGN are above $10^{45} \rm ~ergs~s^{-1}$ ($\sim 2 \times 10^{11} L_\odot$),
and they generally lie between z of 0.3 and 2.5.
We use SED decomposition from the optical to the FIR to 
estimate the AGN luminosities, SFRs and stellar masses of the AGN hosts.

We summarize the results from this study as follows.
\begin{enumerate}

  \item  About 50\% (107 out of 205) of the Type-1 AGNs 
in our sample are individually detected by \herschel. Among these 
AGNs, 68\% show  high levels of star formation (the star formation 
activity contributes over 50\% in the FIR).
\herschel\/ non-detected AGNs were studied using stacking analysis.
On average, they have a similar
level of AGN luminosity and similar optical colors, but the average star formation
activity is several times lower  compared with AGNs individually detected by \herschel.

\item Similarly, about 65\% (55 out of 85) of the Type-2 AGNs are individually detected by \herschel. However,
these objects tend to be at relatively low redshift and some of the detections are a result of 
vigorous star formation, not nuclear activity. We have defined a sample of 15 Type-2 AGN with
properties (M$_{BH}$, Eddington ratio, and redshift) that make them directly comparable with
the Type-1 sample. 

  \item  The FIR-detected Type-1 AGNs and matching Type-2 ones reside in massive galaxies ($\sim 1-2\times 10^{11} ~\rm  M_\odot$).  They harbor supermassive black holes of $\sim  3 \times 10^{8} ~\rm  M_\odot$, 
which accrete at $\sim 10\%$ of the Eddington luminosity.


  \item A warm excess in the MIR was found for eight Type-1 AGNs compared with a local quasar template. 
	This warm excess can be prominent at higher redshifts but is not seen in low redshift quasars.  
        It is not clear whether it changes due to evolution, 
	or whether the warm excess is confined to very luminous quasars. 

\item The 24 $\mu$m-selected sample of Type-1 AGNs includes about twice as many objects 
  as are identified through the SDSS, including the majority of the SDSS identifications. 
   The additional objects have redder optical colors than typical SDSS quasars, due to reddening 
   or intrinsically red quasar continua.

\item As also found, e.g., by Hainline et al. (2013), the strength of the [OIII]$\lambda$5007 line 
increases more rapidly than proportionately to bolometric AGN luminosity. At relatively high redshift (and 
hence high AGN luminosity), detection of [OIII] 
emission from parts of the host galaxy within the spectrograph fiber may contribute to this effect. 
 
\end{enumerate}

These results are discussed further in Xu et al. (2015). 

\section{Acknowledgements}

We thank Xiaohui Fan, Desika Narayanan, and Dan Stark for helpful discussions. 
Marianne Vestergaard provided template spectra and assisted us in spectral line fitting for black hole mass estimation. 
We also thank Yong Shi for communicating results on quasar aromatic band measurements in advance of publication. 
This work is based in part on observations made with {\it Herschel}, a European Space Agency Cornerstone Mission with significant participation by NASA. Additional observations were obtained with {\it Spitzer}, operated by JPL/Caltech. We acknowledge NASA funding for this project through an award for research with Herschel issued by JPL/Caltech. CPH was funded by CONICYT Anillo project ACT-1122. GPS acknowledges support from the Royal Society. Additional support was provided through contract 1255094 from JPL/Caltech to the University of Arizona. This paper also is based in part on work supported by the National Science Foundation under Grant No. 1211349.

Funding for the SDSS and SDSS-II has been provided by the Alfred P. Sloan Foundation, the Participating Institutions, the National Science Foundation, the U.S. Department of Energy, the National Aeronautics and Space Administration, the Japanese Monbukagakusho, the Max Planck Society, and the Higher Education Funding Council for England. The SDSS Web Site is http://www.sdss.org/.

The SDSS is managed by the Astrophysical Research Consortium for the Participating Institutions. The Participating Institutions are the American Museum of Natural History, Astrophysical Institute Potsdam, University of Basel, University of Cambridge, Case Western Reserve University, University of Chicago, Drexel University, Fermilab, the Institute for Advanced Study, the Japan Participation Group, Johns Hopkins University, the Joint Institute for Nuclear Astrophysics, the Kavli Institute for Particle Astrophysics and Cosmology, the Korean Scientist Group, the Chinese Academy of Sciences (LAMOST), Los Alamos National Laboratory, the Max-Planck-Institute for Astronomy (MPIA), the Max-Planck-Institute for Astrophysics (MPA), New Mexico State University, Ohio State University, University of Pittsburgh, University of Portsmouth, Princeton University, the United States Naval Observatory, and the University of Washington.

This publication makes use of data products from the Wide-field Infrared Survey Explorer, which is a joint project of the University of California, Los Angeles, and the Jet Propulsion Laboratory/California Institute of Technology, funded by the National Aeronautics and Space Administration.


\clearpage

\clearpage

\begin{deluxetable}{rrrr}
\tablewidth{0pt}
\tabletypesize{\small}
\tablecaption{\label{cluster} Cluster fields} 
\tablehead{\colhead{Cluster} &   \colhead{RA(J2000)} & \colhead{Dec.(J2000)} & \colhead{Redshift} }

\startdata
      A68    &        00:37:05.28    &     +09:09:10.8    &       0.255   \\
     A115    &        00:55:50.65    &     +26:24:38.7    &       0.197   \\
     Z348$^a$    &        01:06:49.50    &     +01:03:22.1    &       0.160   \\
     A209    &        01:31:53.00    &     -13:36:34.0    &       0.206   \\
  RXJ0142    &        01:42:02.64    &     +21:31:19.2    &       0.280   \\
     A267$^a$    &        01:52:48.72    &     +01:01:08.4    &       0.230   \\
     A291    &        02:01:43.11    &     -02:11:48.1    &       0.196   \\
     A383$^a$    &        02:48:02.00    &     -03:32:15.0    &       0.188   \\
\enddata
\tablenotetext{a}{Type 2 AGNs are selected using MMT/Hectospec spectra in these
21 cluster fields. Because the selection depends on emission line ratios, 
we did not select Type 2 AGNs in 9 cluster fields where 
the spectra were not flux calibrated due to the lack of standard stars. 
}
\end{deluxetable}

\begin{deluxetable}{lcccccc}
\tablewidth{0pt}
\tablecaption{Herschel Photometry parameters \label{tab:herschel_photo}}
\tablehead{
\colhead{} & \colhead{units} & \colhead{100\,$\mu$m} & \colhead{160\,$\mu$m} & \colhead{250\,$\mu$m} & \colhead{350\,$\mu$m} & \colhead{500\,$\mu$m}} 

\startdata
pixel size & arcsec  & 3 & 3  & 6 & 9 & 12 \\
FWHM  & arcsec & 6.8 & 11.4  & 18.1 & 24.8 & 36.6\\
Photometry radius &arcsec  & 6 & 12  &22 &  27 &36 \\
aperture correction & arcsec & 1.706 & 1.499  & 1.229 & 1.120 & 1.211\\
sky annulus & arcsec&  N/A & N/A  &24-60  & 36-90  & 48-120\\ 
\enddata

\end{deluxetable}

\begin{landscape}
\begin{deluxetable}{lrrcr rrrrr rrrrr}
\tablewidth{0pt}
 \tabletypesize{\scriptsize} 
\tablecaption{\label{table_fluxes_type1} {Fluxes of the 24 $\mu$m-selected  \herschel-detected Type-1 AGN sample}\tablenotemark{a}}

\tablehead{ \colhead{\#} &\colhead{Source} &   \colhead{R.A.} & \colhead{Dec.} &    \colhead{J} & \colhead{K} & \colhead{3.4$\mu$m} & \colhead{4.6$\mu$m} &  \colhead{$12\mu$m} & \colhead{$24\mu$m} & \colhead{$100\mu$m} & \colhead{$160\mu$m} &  \colhead{$250\mu$m} & \colhead{$350\mu$m} & \colhead{$500\mu$m} \\
\colhead{ } &\colhead{LIRAS} &  \colhead{J2000} & \colhead{J2000} &    \colhead{(mJy)} & \colhead{(mJy)} & \colhead{(mJy)} & \colhead{(mJy)} & \colhead{(mJy)} & \colhead{(mJy)} & \colhead{(mJy)} & \colhead{(mJy)} & \colhead{(mJy)} & \colhead{(mJy)} & \colhead{(mJy)}  }
 \startdata
  1    &                 J024818.61-031956.9    &      42.0775358    &   -3.3324640    &       1.131    &    1.784    &        1.52    &     1.54    &     4.26    &     8.27    &        42.58    &        60.40    &        53.94    &        35.84    &        24.08   \\
  2    &                 J020120.00-022447.7    &      30.3333465    &   -2.4132541    &       1.361    &    2.683    &        2.63    &     3.57    &    10.36    &    18.43    &        98.39    &        49.37    &        49.65    &        14.15    &       $<$12.39   \\
  3    &                 J212928.91+000415.9    &     322.3704400    &    0.0710740    &       0.518    &    0.833    &        0.46    &     0.44    &     1.09    &     2.56    &        24.91    &        42.34    &        26.08    &       $<$10.99    &       $<$12.39   \\
  4    &                 J073243.17+314111.4    &     113.1798658    &   31.6865098    &       0.279    &    0.527    &        0.50    &     0.54    &     1.32    &     3.05    &       $<$10.89    &        24.61    &        15.24    &        14.19    &       $<$12.39   \\
  5    &                 J082626.42+041231.0    &     126.6100938    &    4.2086221    &       0.143    &    0.300    &        0.25    &     0.25    &     0.56    &     1.37    &        11.40    &       $<$16.89    &       $<$10.19    &       $<$10.99    &       $<$12.39   \\
  6    &                 J084352.28+292854.0    &     130.9678493    &   29.4819339    &       0.378    &    0.686    &        0.42    &     0.36    &     0.80    &     1.33    &       $<$10.89    &       $<$16.89    &        11.68    &       $<$10.99    &       $<$12.39   \\
  7    &                 J014200.86+213752.6    &      25.5035744    &   21.6312836    &       0.095    &    0.186    &        0.16    &     0.18    &     0.46    &     1.13    &        19.02    &        32.18    &        13.94    &       $<$10.99    &       $<$12.39   \\
\enddata

\tablenotetext{a}{Upper limits are 3\,$\sigma$. }
\end{deluxetable}
\end{landscape}

\begin{deluxetable}{lrrrcr rrrrr}
\rotate
\tablewidth{0pt}
\tabletypesize{\footnotesize}
\tablecaption{\label{table_fluxes} {Fluxes of 24 $\mu$m-selected  \herschel-non-detected Type-1 AGN sample}}

\tablehead{\colhead{\#}& \colhead{Source} &  \colhead{R.A.} & \colhead{Dec.} &   \colhead{z} & \colhead{J} & \colhead{K} & \colhead{3.4$\mu$m} & \colhead{4.6$\mu$m} &  \colhead{$12\mu$m} & \colhead{$24\mu$m}  \\
 \colhead{ } & \colhead{LIRAS} &  \colhead{J2000} & \colhead{J2000} &   \colhead{ } & \colhead{(mJy)} & \colhead{(mJy)} & \colhead{(mJy)} & \colhead{(mJy)} & \colhead{(mJy)} & \colhead{(mJy)}  }
 \startdata
  1            &        J145617.60+222124.9    &     224.0733159    &   22.3569112    &    0.111 &       0.371    &    0.446    &        0.26    &     0.24    &     0.59    &     1.00      \\
  2            &        J212932.89+001045.4    &     322.3870342    &    0.1792855    &    0.132 &       0.460    &    0.820    &        0.63    &     0.68    &     1.09    &     2.42      \\
  3            &        J145735.11+223202.4    &     224.3963065    &   22.5339901    &    0.149 &       0.083    &    0.096    &        0.10    &     0.45    &     4.62    &    19.87      \\
  4            &        J020218.53-020243.4    &      30.5772248    &   -2.0453957    &    0.249 &       0.068    &    0.111    &        0.17    &     0.25    &     0.39    &     1.58      \\
  5            &        J140127.69+025606.2    &     210.3653800    &    2.9350470    &    0.265 &       0.175    &    0.313    &        0.79    &     1.30    &     3.30    &     4.68      \\
  6            &        J145720.55+223239.1    &     224.3356237    &   22.5442076    &    0.289 &       0.055    &    0.094    &        0.10    &     0.13    &     0.55    &     1.21      \\
  7            &        J024851.43-032249.3    &      42.2143100    &   -3.3803662    &    0.300 &       0.144    &    0.349    &        0.60    &     0.87    &     1.26    &     2.19      \\
  8            &        J140019.72+030346.1    &     210.0821561    &    3.0628171    &    0.319 &       0.245    &    0.431    &        0.40    &     0.42    &     0.72    &     1.72      \\
\enddata

\end{deluxetable}

\begin{landscape}
\begin{deluxetable}{llrrrrrrrrrrrrr}
\tablewidth{0pt}
\tabletypesize{\footnotesize}
\tablecaption{\label{table_luminosity_type1} Redshifts, luminosities, broad emission line widths, black hole masses, and 
stellar masses for the {\it Herschel}-detected Type-1 sources}

\tablehead{ \colhead{\#} & \colhead{Source} &  \colhead{z} &  \colhead{$L_{\rm 0.5-8~keV}$} & \colhead{$L_{\rm B}$} & \colhead{$L_{\rm tot,IR}$} & \colhead{$L_{\rm SF,IR}$\tablenotemark{a}} &  \colhead{$L_{\rm AGN,IR}$} & \colhead{$L_{\rm warm}$\tablenotemark{b}} & \colhead{Line} &  \colhead{FWHM} &\colhead{$M_\bullet$\tablenotemark{c} } &  \colhead{$M_*$}  & \colhead{$$} \\
\colhead{} & \colhead{LIRAS} & \colhead{} & \colhead{ $10^{10}  \rm L_\odot$} & \colhead{$10^{11}  \rm L_\odot$} &   \colhead{$10^{11}  \rm L_\odot$} & \colhead{$10^{11}  \rm L_\odot$} & \colhead{$10^{11}  \rm L_\odot$} & \colhead{$10^{11}  \rm L_\odot$} & \colhead{} & \colhead{$\rm km~s^{-1}$}& \colhead{$10^8  \rm M_\odot$} & \colhead{$10^{11}  \rm M_\odot$}  & \colhead{}   }
 \startdata
   1    &            J024818.61-031956.9    &    0.127 &               &     0.71    &     0.41    &        0.22(0.00)&    0.19    &       (0.00)&             H$\beta$      &   2521    &     0.29    &          0.95\tablenotemark{d}   \\
   2    &            J020120.00-022447.7    &    0.136 &               &     1.02    &     0.90    &        0.40(-0.10)&    0.50    &        (0.20)&             H$\beta$      &   2594    &     0.51    &          0.85\tablenotemark{d}   \\
   3    &            J212928.91+000415.9    &    0.180 &          0.05 &     0.01    &     0.71    &        0.61(-0.55)&    0.11    &        (0.09)&             H$\alpha$     &  $>$1200   &    N/A    &          0.82\tablenotemark{d}   \\
   4    &            J073243.17+314111.4    &    0.276 &          0.49 &     1.36    &     0.78    &        0.43(0.00)&    0.36    &       (0.00)&             H$\beta$      &   3313    &     0.69    &          0.87\tablenotemark{d}   \\
   5    &            J082626.42+041231.0    &    0.324 &          0.30 &     0.00    &     0.82    &        0.62(-0.16)&    0.20    &        (0.07)&             H$\beta$      &   3017    &     0.43    &          0.81\tablenotemark{d}   \\
   6    &            J084352.28+292854.0    &    0.336 &               &     1.50    &     0.68    &        0.44(-0.06)&    0.24    &        (0.07)&             H$\beta$      &  11620    &     6.99    &          1.92\tablenotemark{d}   \\
   7    &            J014200.86+213752.6    &    0.381 &               &     1.28    &     1.29    &        1.07(-0.04)&    0.23    &        (0.12)&             H$\beta$      &   2225    &     0.25    &          0.65\tablenotemark{d}   \\
   8    &            J212930.74+001347.2    &    0.395 &               &     2.48    &     7.34    &        1.92(-1.00)&    5.41    &        (6.26)&             H$\beta$      &   2326    &     1.34    &          0.94                                \\
\enddata

\tablenotetext{a}{Numbers in parenthesis are the decrease in the fraction of the SF IR luminosities if we add the warm component as a 
degree of freedom in the SED decomposition precedure. 
For the case that the warm component is required 
in order to obtain good SED fits, we do not provide this  decrease fraction with parenthesis}
\tablenotetext{b}{Numbers in parenthesis are the warm component luminosities if we add the warm component as a 
degree of freedom in the SED decomposition precedure.
For some sources, the warm component is required 
in order to obtain good SED fits. The  warm component luminosities for such sources are provided with no parenthesis.}

\tablenotetext{c}{Some AGNs are broad absorption line quasars, and their broad emission line FWHM is above 1200 
\kms, but is not well constrained. Some AGNs have low S/N optical spectra, and the emission line fitting is 
not good enough for black hole mass estimate.}

\tablenotetext{d}{Host galaxy stellar masses derived from K band luminosity; all other masses are  estimated from the local $M_{\bullet}/M_*$ ratio.}

\end{deluxetable}

\end{landscape}

\begin{deluxetable}{lrrcr rrrrr rrrrr}
\rotate
\tablewidth{0pt}
\tabletypesize{\tiny}
\tablecaption{\label{table_fluxes} {Fluxes of 24 $\mu$m-selected  \herschel-detected Type-2 AGN sample}\tablenotemark{a}}

\tablehead{ \colhead{\#} &\colhead{Source} &   \colhead{R.A.} & \colhead{Dec.} &    \colhead{J} & \colhead{K} & \colhead{3.4$\mu$m} & \colhead{4.6$\mu$m} &  \colhead{$12\mu$m} & \colhead{$24\mu$m} & \colhead{$100\mu$m} & \colhead{$160\mu$m} &  \colhead{$250\mu$m} & \colhead{$350\mu$m} & \colhead{$500\mu$m} \\
\colhead{ } &\colhead{LIRAS} &  \colhead{J2000} & \colhead{J2000} &    \colhead{(mJy)} & \colhead{(mJy)} & \colhead{(mJy)} & \colhead{(mJy)} & \colhead{(mJy)} & \colhead{(mJy)} & \colhead{(mJy)} & \colhead{(mJy)} & \colhead{(mJy)} & \colhead{(mJy)} & \colhead{(mJy)}  }
 \startdata
  1    &                 J101756.90+390528.0    &     154.4870647    &   39.0911158    &       2.140    &    2.337    &        1.14    &     0.86    &     3.33    &     6.41    &           95.45    &          100.84    &           52.66    &           24.71    &              12.98   \\
  2    &                 J101805.64+385009.7    &     154.5235201    &   38.8360315    &       3.217    &    3.689    &        0.00    &     0.00    &     0.00    &     6.15    &                    &                    &          132.04    &           55.62    &              17.77   \\
  3    &                 J145753.24+222422.7    &     224.4718447    &   22.4063113    &       1.190    &    1.550    &        0.70    &     0.53    &     1.01    &     1.51    &           31.49    &           43.06    &           24.72    &           13.63    &           $<$12.39   \\
  4    &                 J133250.99+501816.1    &     203.2124640    &   50.3044596    &       0.445    &    0.801    &        0.50    &     0.45    &     1.12    &     1.49    &           39.55    &           56.07    &           27.41    &           10.88    &           $<$12.39   \\
  5    &                 J083713.49+150037.4    &     129.3061885    &   15.0103981    &       0.681    &    1.135    &        0.54    &     0.40    &     1.00    &     1.23    &                    &                    &           43.56    &           16.00    &           $<$12.39   \\
  6    &                 J080102.09+355132.2    &     120.2587258    &   35.8589486    &       0.432    &    0.519    &        0.32    &     0.25    &     1.00    &     4.93    &           21.28    &           13.53    &            8.82    &        $<$10.99    &           $<$12.39   \\
\enddata

\tablenotetext{a}{Upper limits are 3\,$\sigma$. }
\end{deluxetable}

\begin{deluxetable}{lrrrcr rrrrr}
\rotate
\tablewidth{0pt}
\tabletypesize{\footnotesize}
\tablecaption{\label{table_fluxes} {Fluxes of 24 $\mu$m-selected  \herschel-non-detected Type-2 AGN sample}}

\tablehead{\colhead{\#} &\colhead{Source} &  \colhead{R.A.} & \colhead{Dec.} &   \colhead{z} & \colhead{J} & \colhead{K} & \colhead{3.4$\mu$m} & \colhead{4.6$\mu$m} &  \colhead{$12\mu$m} & \colhead{$24\mu$m}  \\
\colhead{ } & \colhead{LIRAS} &  \colhead{J2000} & \colhead{J2000} &   \colhead{ } & \colhead{(mJy)} & \colhead{(mJy)} & \colhead{(mJy)} & \colhead{(mJy)} & \colhead{(mJy)} & \colhead{(mJy)}  }
 \startdata
  1    &                 J101813.08+390220.2    &     154.5544848    &   39.0389521    &    0.293 &       0.142    &    0.207    &        0.00    &     0.00    &     0.00    &     1.27   \\
  2    &                 J083805.70+145152.9    &     129.5237579    &   14.8646849    &    0.345 &       0.097    &    0.263    &        0.79    &     1.30    &     3.30    &     9.08   \\
  3    &                 J073301.63+314042.2    &     113.2567957    &   31.6783963    &    0.115 &       0.525    &    0.653    &        0.37    &     0.38    &     1.59    &     6.23   \\
  4    &                 J024852.78-033551.5    &      42.2199003    &   -3.5976449    &    0.137 &       0.882    &    1.177    &        0.64    &     0.51    &     0.69    &     1.63   \\
  5    &                 J101814.47+385217.0    &     154.5602787    &   38.8716572    &    0.147 &       0.160    &    0.199    &        0.12    &     0.07    &     0.53    &     1.52   \\
  6    &                 J164119.41+462953.5    &     250.3308715    &   46.4981864    &    0.191 &       0.553    &    0.779    &        0.57    &     0.62    &     1.37    &     2.99   \\
  7    &                 J010623.62+005143.0    &      16.5984100    &    0.8622204    &    0.195 &       0.725    &    1.022    &        0.54    &     0.36    &     0.55    &     1.73   \\
  8    &                 J090048.78+204448.2    &     135.2032534    &   20.7467242    &    0.329 &       0.032    &    0.053    &        0.00    &     0.00    &     0.00    &     1.48   \\
\enddata

\end{deluxetable}

\begin{deluxetable}{llrrrrrrrr}
\tablewidth{0pt}
\tabletypesize{\footnotesize}
\tablecaption{\label{table_luminosity} Redshifts and derived parameters for the {\it Herschel}-detected Type-2 sources}

\tablehead{ \colhead{\#} & \colhead{Source} &  \colhead{z} &   \colhead{$L_{\rm tot,IR}$} & \colhead{$L_{\rm SB,IR}$} &  \colhead{$L_{\rm AGN,IR}$} & \colhead{$L_{\rm AGN,total}$}  & \colhead{M*} & \colhead{Type$^d$} \\
\colhead{} & \colhead{LIRAS} & \colhead{} &  \colhead{$10^{11}  \rm L_\odot$} &   \colhead{$10^{11}  \rm L_\odot$} & \colhead{$10^{11}  \rm L_\odot$} & \colhead{$10^{11}  \rm L_\odot$}  & \colhead{$10^{11} \rm M_\odot$} }

 \startdata
   1    &            J101756.90+390528.0    &    0.054 &        0.10    &     0.09    &     0.01    &     0.03    & 0.33 &  S  \\
   2    &            J101805.64+385009.7    &    0.067 &        0.21    &     0.21    &     0.00    &     0.03    &  0.77 & S0/E \\
   3    &            J145753.24+222422.7    &    0.109 &        0.13    &     0.13    &     0.00    &     0.01    &  0.76 & C \\
   4    &            J133250.99+501816.1    &    0.110 &        0.12    &     0.11    &     0.01    &     0.32    &  0.31 & S0/E \\
   5    &            J083713.49+150037.4    &    0.141 &        0.23    &     0.22    &     0.01    &     0.13    &  0.80 &  E \\
   6    &            J080102.09+355132.2$^{b}$  &    0.160 &        0.34    &     0.19    &     0.15    &     2.14    & 0.60 & E  \\
   7    &            J133644.33+405854.9  &    0.169 &        0.30    &     0.23    &     0.08    &     0.32    &   1.00 & S \\
   8    &            J101623.97+385840.1    &    0.169 &        4.75    &     4.72    &     0.03    &     0.70    &   0.68 & E  \\
\enddata
\tablenotetext{a}{High Luminosity Subsample (HLS)}


\tablenotetext{b}{AGN-dominated at 24 $\mu$m}

\tablenotetext{c}{Comparison sample (see text)}

\tablenotetext{d} {S = spiral, E = elliptical, C = too compact to classify, I = interacting}


\end{deluxetable}

\clearpage

\begin{figure}[!hbt]
\centering
\includegraphics[width=0.5\textwidth,angle=00]{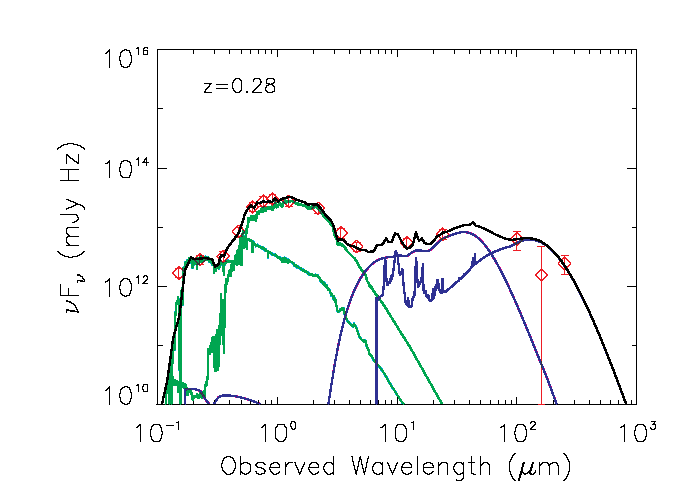}
\includegraphics[width=0.5\textwidth,angle=00]{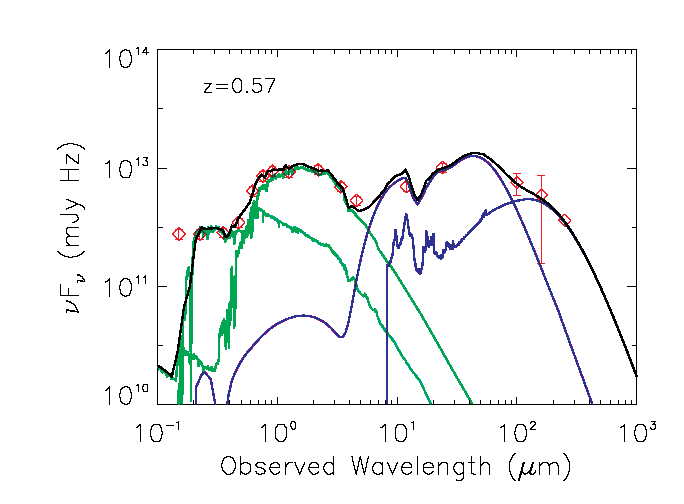}
\caption{The average SEDs for Type-2 AGNs with no formal FIR detections in our sample
in two discrete redshift bins: 
$z=$0.0-0.4 (upper, 14 sources), and
0.4-0.8 (lower, 13 sources). Symbols and line colors are the same as in Figure \ref{fig:dcp_example}.
}
\label{fig:stack_sed}
\end{figure}

\begin{figure}[!hbt]
\centering
\includegraphics[width=0.8\textwidth]{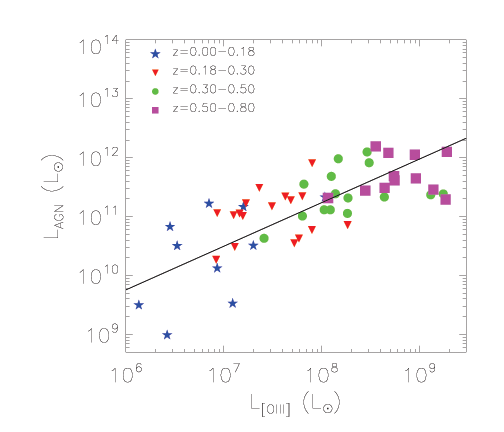}
\caption{
The relation between  AGN total luminosity and
 $\mbox{\rm \oiii} ~\lambda$5007 luminosity for \herschel-detected Type-2 AGNs in
our sample. The fitted line (all points with equal weights) has a slope of 0.74.
}
\label{fig:l5007_lir_all}
\end{figure}

\begin{figure}[!hbt]
\centering
\includegraphics[width=0.6\textwidth,angle =90]{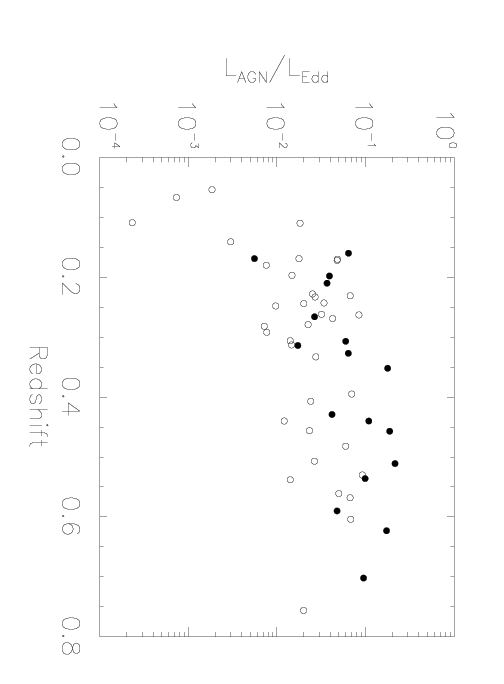}
\caption{The ratio of AGN luminosity to Eddington luminosity for our Type-2 AGNs. Filled circles are for
AGN-dominated sources, while open ones are for SF-dominated.}
\label{fig:stellar_blackhole_eddington}
\end{figure}

\begin{figure}[!hbt]
\centering
\includegraphics[width=0.6\textwidth,angle=90]{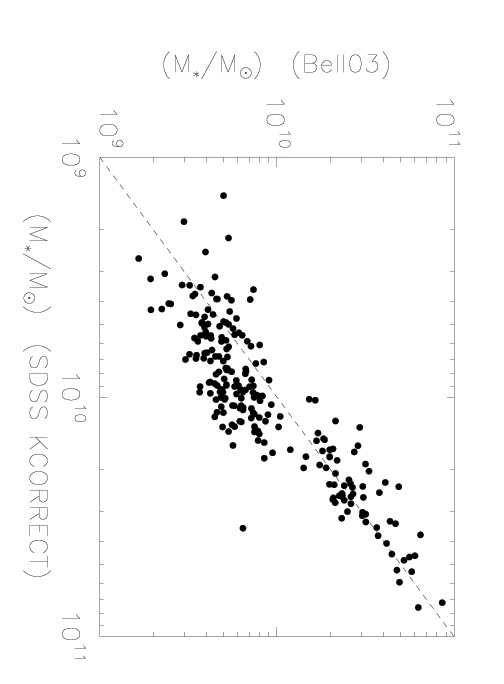}
\caption{Comparison of the stellar mass calculated from \citet{bell03} and SDSS KCORRECT
 (private communication, Krystal Tyler). The SDSS KCORRECT stellar masses are 
based on the Bruzual-Charlot stellar evolution synthesis.
The stellar masses derived from \citet{bell03} are consistent with those from SDSS KCORRECT.
 }
\label{fig:stellarmass_comparison}
\end{figure}

\begin{figure}[!hbt]
\centering
\includegraphics[width=0.6\textwidth, angle = 90]{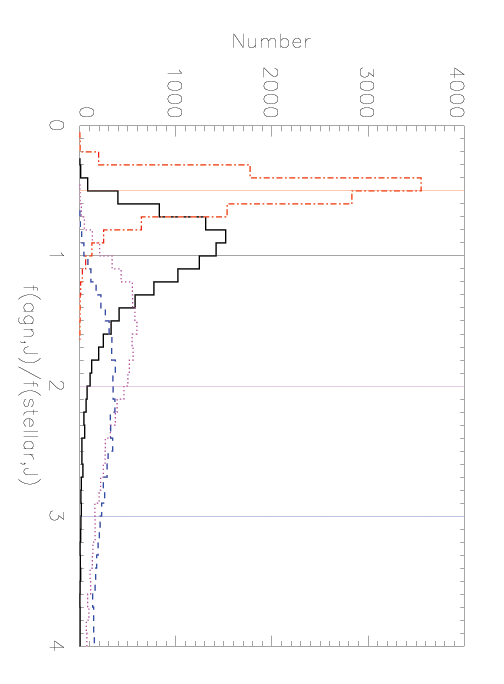}
\caption{Simulation results for the ability of SED decomposition to constrain the stellar 
component in the NIR. The simulation is performed 10000 times
for each fixed value of AGN and stellar component in the rest-frame J-band. 
The derived results for the input flux ratio $\rm flux_{AGN,J}/flux_{Stellar,J} =$ 0.5, 1, 2, and 3, 
are shown in red dash-dot, black solid, magenta dotted, and blue dashed lines. 
}
\label{fig:simulation_ratio_agnstar}
\end{figure}

\begin{figure}[!hbt]
\centering
\includegraphics[width=0.5\textwidth,angle=90]{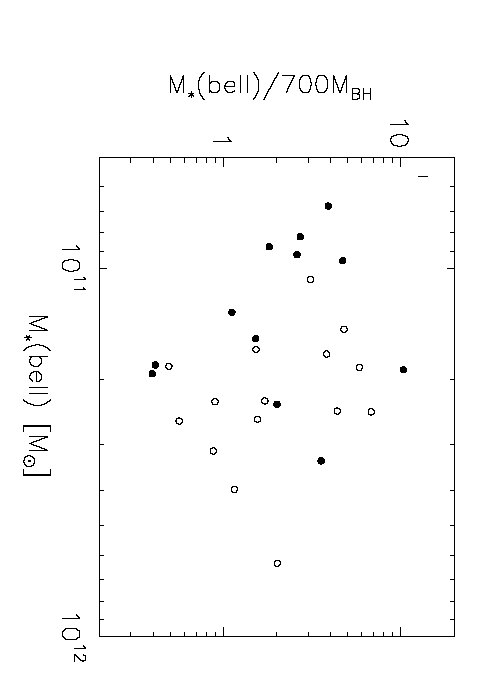}
\caption{Comparison of the stellar masses 
estimated by K-band luminosity using the equation from \citet{bell03} (See Section 6.4)
and the stellar masses derived from the local mass ratio $M_*/M_{\bullet}= 700$. 
The filled and unfilled  circles are AGNs at $z< 0.6$ and $0.6<z<1.2$, respectively. 
}
\label{fig:stellar_mass_more}
\end{figure}

\clearpage
\appendix

\section{ Images of Type-2 Host Galaxies}

\begin{figure}[!hbt]
\centering
\includegraphics[width=0.42\textwidth, angle = 00]{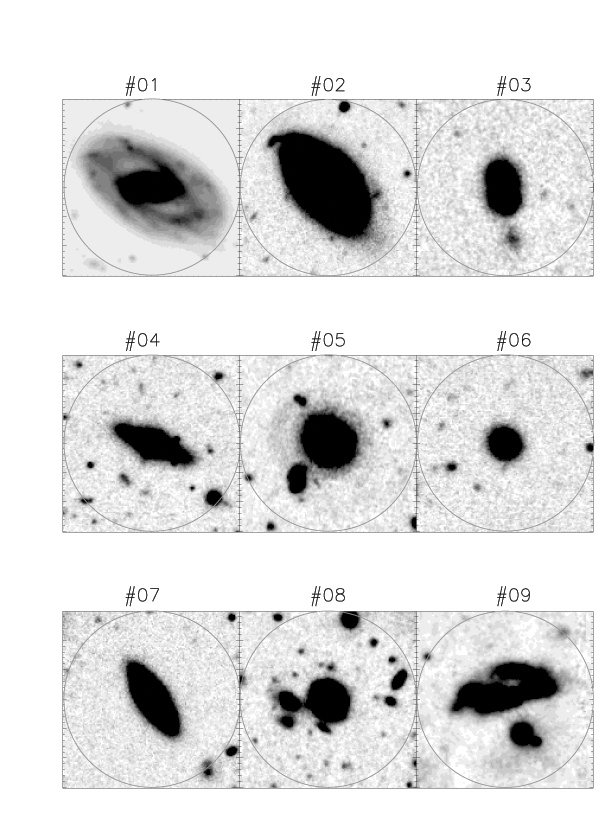}
\includegraphics[width=0.42\textwidth, angle = 00]{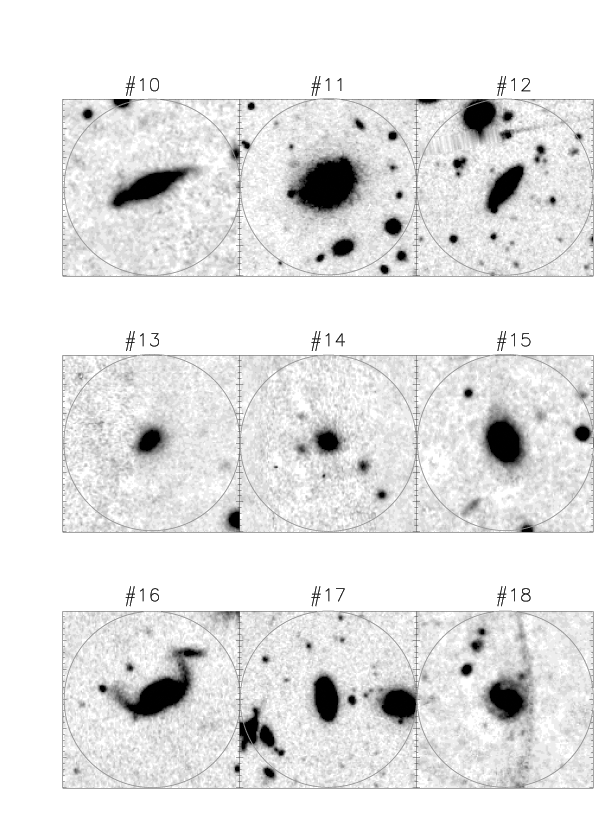}
\includegraphics[width=0.42\textwidth, angle = 00]{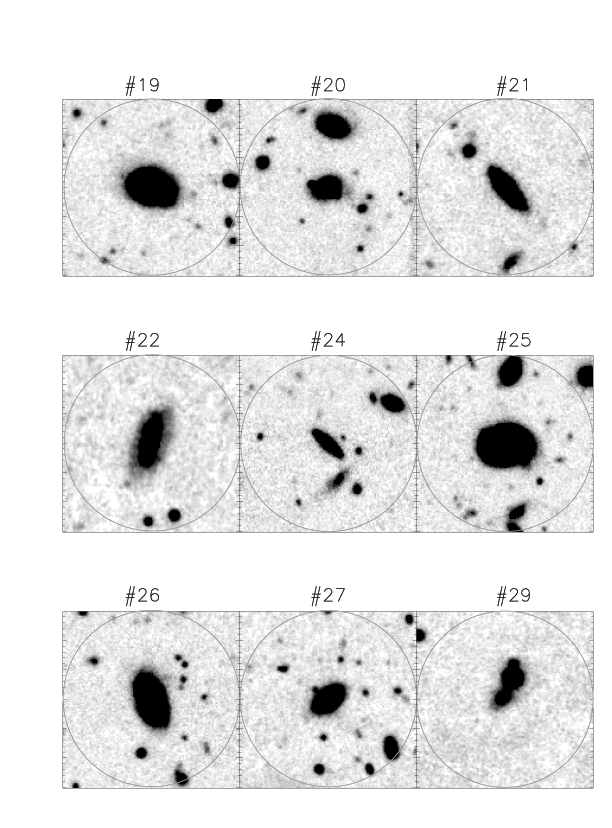}
\includegraphics[width=0.42\textwidth, angle = 00]{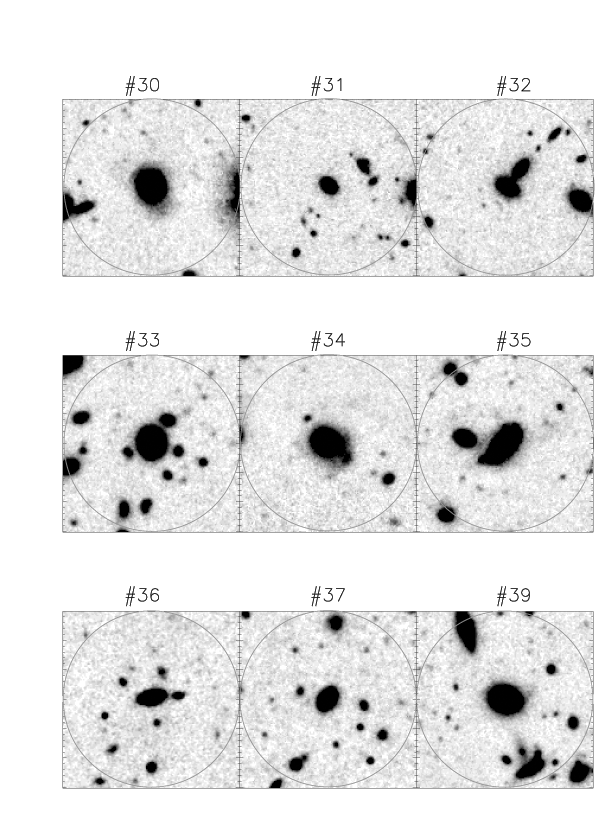}
\caption{Subaru images of  \herschel-detected  Type-2 AGNs. The circle radius is 15\arcsec.
The 1st part of a continued figure. 
}
\label{fig:subaru1}
\end{figure}

\setcounter{figure}{19}

\begin{figure}[!hbt]
\centering
\includegraphics[width=0.42\textwidth, angle = 00]{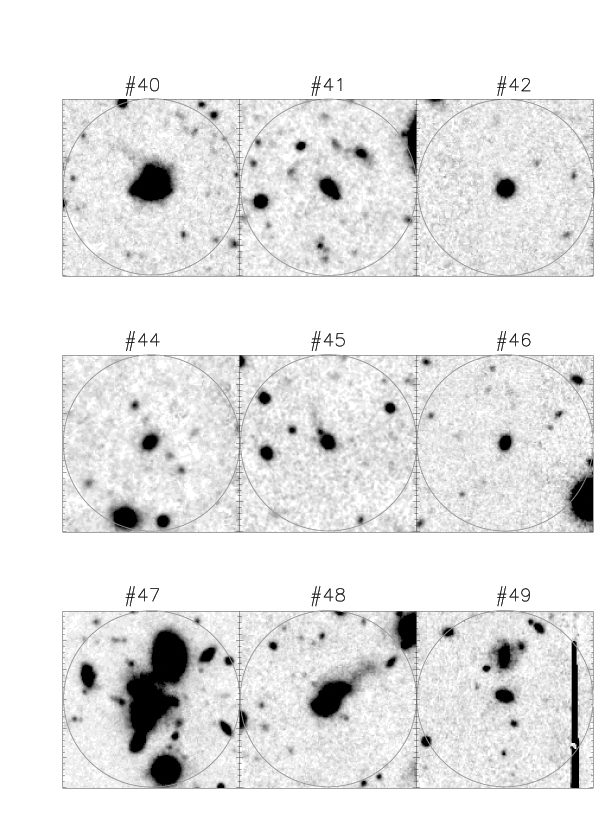}
\includegraphics[width=0.42\textwidth, angle = 00]{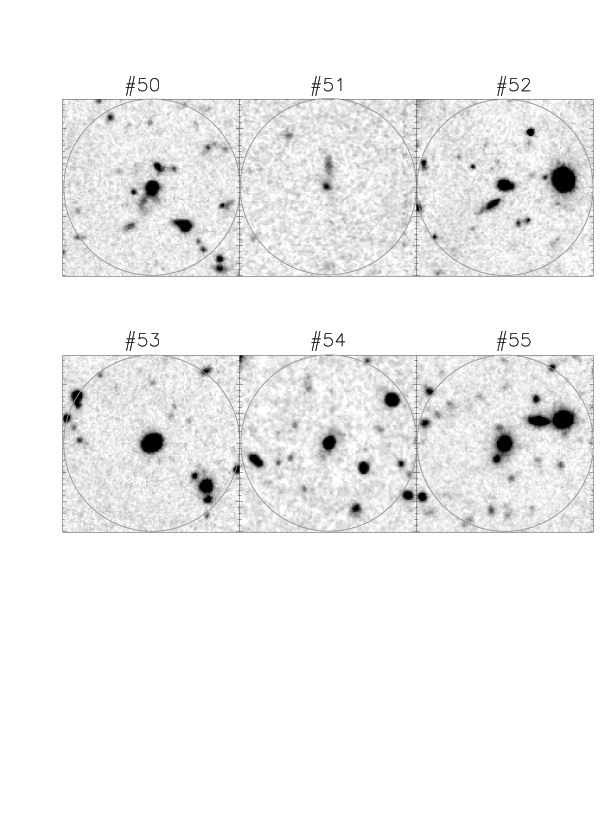}
\caption{Subaru images of  \herschel-detected  Type-2 AGNs. The circle radius is 15\arcsec.
The 2nd part of a continued figure. 
}
\label{fig:subaru5}
\end{figure}

\section{24 $\mu$m-Selected Type-1 AGN Sample in the LoCuSS Fields}\label{coverage}

In total, we detected 2439  sources  with 24 $\mu$m flux above 1 mJy
 in the LoCuSS fields. 
Out of these, the following 541 sources were
  not included in the target list for the Hectospec spectroscopic follow-up:
  \begin{enumerate}
    \item 71 sources that were outside the available near-infrared images. 
    \item 168 sources that were identified as stars.
    \item 373 sources with no obvious optical/near-infrared counterparts (The 5-$\sigma$ detection limit of the Subaru images in r or i band is $\sim$ 25 magnitude and the 5-$\sigma$ detection limit at K-band is 19 mag (Vega)).).
  \end{enumerate}
The remaining 1827 24 $\mu$m sources are
likely to be extragalactic. We may have discarded a number of extragalactic sources with faint
optical/near-infrared counterparts (category 3 above) although some
fraction of the category 3 sources is expected to be asteroids.

Among these 1827 sources, 1729 were observed by Hectospec while another 18 sources
have spectroscopic information from SDSS.  The completeness
of the spectroscopic coverage is therefore about 94.6\%.  
Among the 1729 sources targeted by Hectospec, 1263 sources have
produced spectroscopic redshifts with the corresponding success rate
of 73\%.  However, the sources that did not produce spectroscopic redshifts
are unlikely to be Type-1 AGNs.  Therefore, our
24 $\mu$m-selected Type-1 AGN sample is expected to be complete at the
$\sim$94\% level, which is the completeness of our spectroscopic
coverage. Thus, we have 205 sources that satisfy our Type-1 AGN selection
 criteria (See Section \ref{sample_type1}), 177 confirmed with Hectospec spectra, and
28 confirmed with SDSS spectra.

\section{Black Hole Mass Estimate\label{black_hole_mass}}

The following methods were used to estimate black hole masses from our spectra:

\begin{enumerate}
\item {\bf FWHM(\mbox{\boldmath \Hbeta}) and
\mbox{\boldmath $L_{\lambda}$(5100\,\AA)}.} For the optical
continuum luminosity and FWHM of the \Hbeta\ broad component,
\begin{equation}
\log \,M_{\rm BH} (\rm H\beta) = 
   \log \,\left[ \left(\frac{\rm FWHM(H\beta)}{1000~km~s^{-1}} \right)^2 ~ 
   \left( \frac{\lambda \it L_{\lambda} {\rm (5100\,\AA)}}{10^{44} \rm erg~s^{-1}}\right)^{0.50} 
        \right] + (6.91 \pm 0.02).
\label{logMopt_L51.eq}
\end{equation}
The sample standard deviation of the weighted average zeropoint offset is $\pm 0.43$ dex
\citep{vestergaard06}.

\item {\bf FWHM(\mbox{\boldmath \mgii})}. 
For a given wavelength,
\lam, the black hole mass based on \mgii{} was obtained according to:
\begin{equation}
M_{\rm BH} = 10^{zp(\lambda)} \left[\frac{{\rm FWHM(Mg{\sc II})}}{1000 \rm \,km~s^{-1}}\right]^2 \left[\frac{\lambda L_{\lambda}}{10^{44}  \rm ~ergs~s^{-1}}\right]^{0.5}
\end{equation}
where $zp(\lambda)$ is 6.72, 6.79, 6.86, and 6.96 for \lam 1350\,\AA,
\lam 2100\,\AA, \lam 3000\,\AA, and \lam 5100\,\AA, respectively.
The 1\,$\sigma$ scatter in the absolute zero-points, $zp$, is 0.55\,dex
\citep{vestergaard09}.

\item {\bf FWHM(\mbox{\boldmath \civ}) and
\mbox{\boldmath $L_{\lambda}$(1350\,\AA)}.} For the ultraviolet
continuum luminosity and the FWHM of the \civ\ line,
\begin{equation}
\log \,M_{\rm BH} \mbox{(\rm \civ)} = 
   \log \,\left[ \left(\rm \frac{FWHM\mbox{(\rm \civ)}}{1000~km~s^{-1}} \right)^2 ~ 
   \left( \frac{\lambda L_{\lambda} (1350\,{\rm \AA})}{10^{44} \rm erg~s^{-
1}}\right)^{0.53} 
        \right] + (6.66 \pm 0.01).
\label{logMuv_fw.eq}
\end{equation}
The sample standard deviation of the weighted average zeropoint offset
is $\pm0.36$ dex \citep{vestergaard06}.
The $L_{\lambda}$(1450\,\AA) luminosity
is equivalent to $L_{\lambda}$(1350\,\AA) in the equation above
without error or penalty in precision \citep{vestergaard06}.

\end{enumerate}

\begin{figure}[!hbt]
\centering
\includegraphics[width=0.35\textwidth, angle=90]{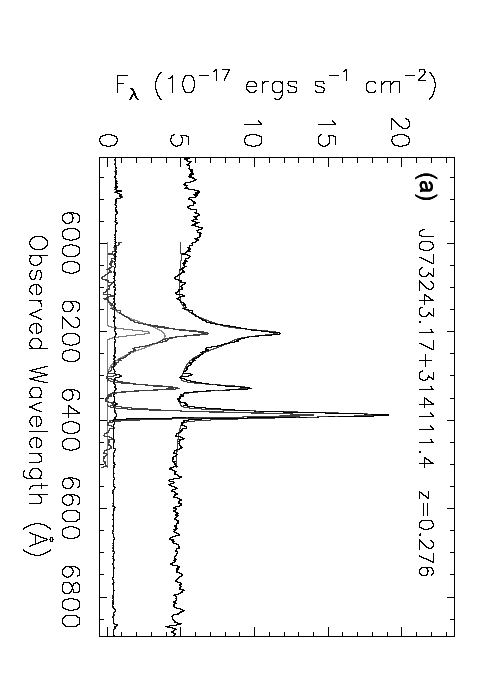}
\includegraphics[width=0.35\textwidth, angle=90]{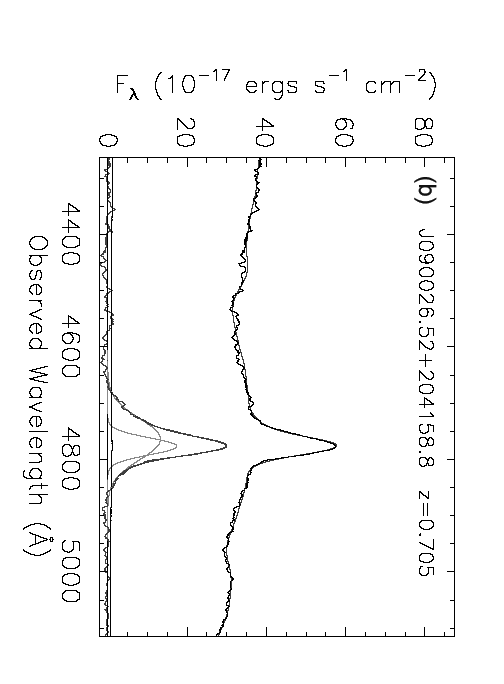}
\includegraphics[width=0.35\textwidth, angle=90]{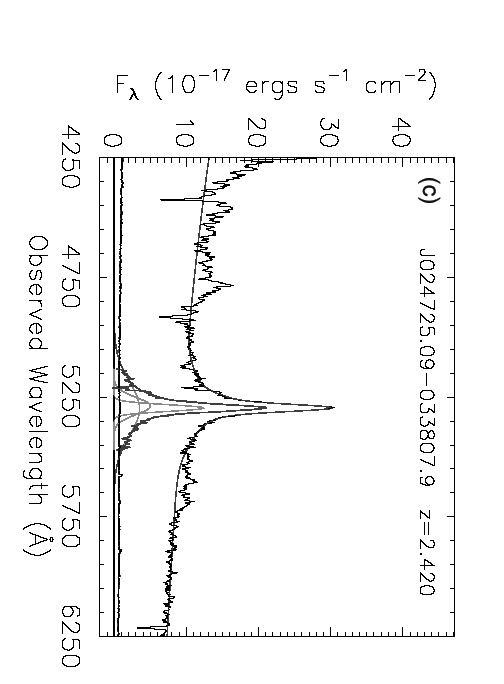}
\caption{Examples of broad emission line fits. 
(a): \Hbeta; (b): \mgii; (c): \civ. 
For each panel, the upper black line  shows the original SED. 
The lower blue line shows the continuum and Fe subtracted SED. 
The upper magenta line shows the full fits;
the lower magenta line shows the  fits for the emission lines;
the gray lines show the flux density errors;
the green lines show the broad Gaussian components, 
while the red lines show the narrow Gaussian components.
}
\label{fig:broad_emission}
\end{figure}

Figure \ref{fig:broad_emission} shows the examples of broad emission line fits for 
\Hbeta, \mgii, and \civ, respectively.

\section{Correction of the AGN Template for Star Formation}

The template we use for the intrinsic AGN SED was built from a detailed set of observations of a representative set of optically selected quasars by \citet{elvis94}. A more recent study by \citet{richards06} used a similar approach and derived a virtually identical template. The excellent agreement is encouraging; for example, our results are independent of which template we use. However, neither study attempted to correct the templates for the far infrared emission due to star formation. Doing so is challenging because one needs an independent, extinction-free estimate of the rate of star formation in the quasar host galaxies. The 11.3 $\mu$m aromatic feature is an appropriate indicator, particularly since it is not strongly affected by an AGN \citep{diamond10}. We have therefore used a large set of measurements of this feature in quasar spectra \citep{shi14}, along with a star forming galaxy far infrared template \citep{rieke09} to estimate the necessary correction. The approach was to correlate the equivalent width of the 11.3 $\mu$m feature with the ratio of fluxes at 25 and 60 $\mu$m (IRAS) or at 24 and 70 $\mu$m (MIPS) to determine the influence of star formation on the far infrared spectrum  in a variety of galaxies with and without AGN. We then used the relation derived from this correlation analysis and the average EW of the 11.3 $\mu$m feature for the quasar sample used by \citet{elvis94} to determine how to adjust their template in the far infrared.  

The initial template we used was for radio-quiet quasars; \citet{elvis94} list 19 of these sources with IRAS detections, and they would have been most influential in determining the far infrared behavior of their template (we return to the IRAS upper limits later). Of those 19, we have 11.3 $\mu$m EW measurements for 15 (79\%), with an average value of 0.037 $\mu$m (standard deviation of the mean = 0.007 $\mu$m). A linear fit to the dependence of EWs vs. infrared flux ratios indicates that the ratio of IRAS 60 to 25 $\mu$m flux densities for the \citet{elvis94} template has been boosted by a factor of 1.24 due to star formation, relative to the case for an EW of 0.0. However, the baseline in EW is small, so we repeated the determination adding the galaxies from \citet{brandl06} (which we selected because the methodology for determining EWs was similar to the method for the quasars). This reference includes cases with EW up to $\sim$ 0.9, thus extending the baseline and improving the determination of the slope of the relation. This fit indicated a star-formation induced boost in the far infrared flux ratio for the Elvis template by a factor of 1.27. When we added the radio loud quasars in the Elvis sample plus additional PG quasars with 11.3 $\mu$m and far infrared measurements, and substituted MIPS for IRAS measurements when they were available, we got a value of 1.27. This last correlation is illustrated in Figure 21. 

\begin{figure}[!hbt]
\centering
\includegraphics[width=0.7\textwidth]{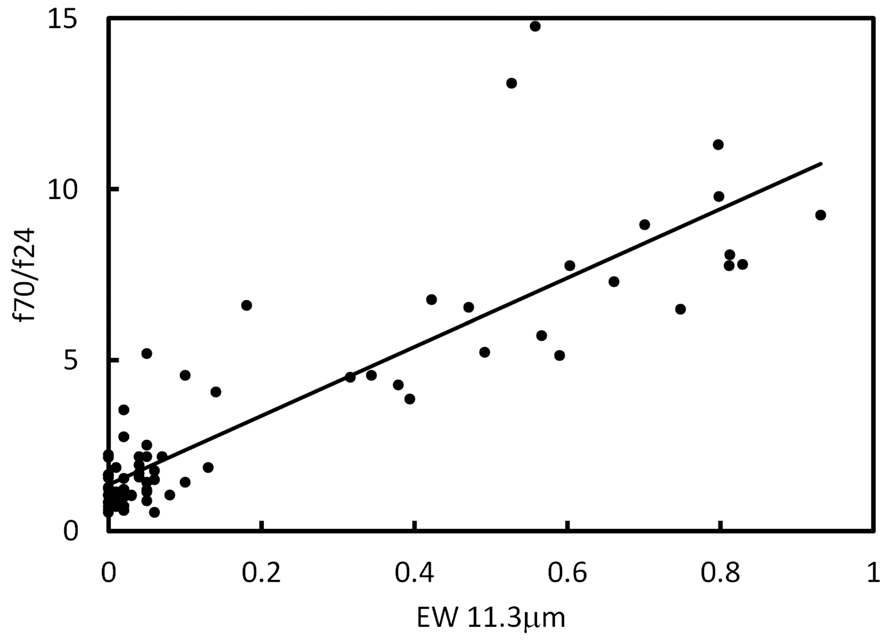}
\caption{The relation between the equivalent width of the 11.3 $\mu$m aromatic feature
and the ratio of flux densities at 70 and 24 $\mu$m (from MIPS) or at 60 and 25 $\mu$m 
(from IRAS, if MIPS measurements are not available). The data are from
\citet{shi14} and \citet{brandl06}
}. 
\label{fig:pah_irslope}
\end{figure}

With a determination of the size of the star-formation boost in the flux ratio, we subtracted a star-forming galaxy template (specifically for L(TIR) = 10$^{11}$ L$_\odot$ \citep{rieke09}) from the Elvis AGN template. We used synthetic photometry on the f60/f25 flux density ratio to match the results from the correlation analysis based on the EW of the 11.3 $\mu$m feature. 

The adjusted AGN template may be an extreme case, since we did not include the galaxies in the sample of \citet{elvis94} for which there were only IRAS upper limits. These galaxies should include those with the weakest star formation relative to the AGN, as well as some that are just fainter then the detected ones at all wavelengths. It is not possible to reconstruct exactly what effect the upper limit cases would have had on the published template, but presumably they tended to make it fainter in the far infrared than it would have been based only on the IRAS detected cases. Thus, we consider our adjusted AGN template to be a limiting case for the maximum plausible far infrared contribution from star formation, and take the unadjusted template to be the limiting case in the other direction.

This approach provides a correction out to 100 $\mu$m (rest). Beyond this wavelength, the Elvis 
template is a power law interpolation to the radio regime. There are very few examples of quasars that 
can be shown to have very low leves of star formation and at the same time have sufficiently sensitive 
measurements of upper limits at wavelengths longer than 100 $\mu$m. Two examples, PG 1501+106 and PG 1411+442, indicate that the power law substantially overestimates the fluxes in this region. 
Therefore, a more realistic replacement is a blackbody of 118K,
with a wavelength dependent emissivity proportional to $\lambda^{-1.5}$ and scaled to match smoothly 
to the corrected SED at wavelengths short of 100 $\mu$m. 

\newpage

\end{document}